\documentclass[aps,pra,notitlepage,twocolumn,amsmath,amssymb,showpacs,nofootinbib]{revtex4-2}
\usepackage{graphicx}
\usepackage{amsmath}
\usepackage{bm}
\usepackage{xcolor}

\newcommand{\be}{\begin{equation}}  
\newcommand{\ee}{\end{equation}}
\newcommand{\ba}{\begin{array}}
\newcommand{\ea}{\end{array}}
\newcommand{\bea}{\begin{eqnarray}}
\newcommand{\eea}{\end{eqnarray}}

\newcommand{\bra}{\langle}
\newcommand{\ket}{\rangle}

\newcommand{\nn}{\nonumber}

\begin{document}

\title{Steady state in strong system-bath coupling regime: mean force Gibbs state versus reaction coordinate}

\author{Camille L. Latune$^{1,2}$} 
\affiliation{$^1$Univ Lyon, ENS de Lyon, CNRS, Laboratoire de Physique,F-69342 Lyon, France\\
$^2$Quantum Research Group, School of Chemistry and Physics, University of KwaZulu-Natal, Durban, KwaZulu-Natal, 4001, South Africa}

\date{\today}
\begin{abstract}
Motivated by the growing importance of strong system-bath coupling in several branches of quantum information and related technological applications, we analyze and compare two strategies currently used to obtain (approximately) steady states in strong coupling regime. The first strategy is based on perturbative expansions while the second one uses reaction coordinate mapping. Focusing on the widely used spin-boson model, we show that the predictions of these two strategies coincide in many situations. This confirms and strengthens the relevance of both techniques. Beyond that, it is also crucial to know precisely their respective range of validity. In that perspective, thanks to their different limitations, we use one to benchmark the other. We introduce and successfully test some very simple validity criteria for both strategies, bringing some answers to the question of the validity range. 
\end{abstract}

\maketitle

\section{Introduction}
It is notoriously challenging to describe the dynamics and the steady states of quantum systems coupled to noisy environment\cite{FrancescoBook}. This is particularly true when the coupling strength is such that system-environment correlations cannot be neglected, invalidating the traditional Born and Markov approximations \cite{CohenBook}, characterizing the so-called strong coupling regime.

 Still, strong coupling effects are playing increasing role in quantum transport \cite{Chin_2012, Ribeiro_2015,Strasberg_2018,Correa_2019,Moreira_2020,Harush_2020,Dwiputra_2021, Anto_2021,Trushechkin_2019}, quantum sensing \cite{Correa_2017, Mehboudi_2019,Mejia_2021}, quantum thermal engines \cite{Gelbwaser_2015,Strasberg_2016,Newman_2017,Perarnau_2018,Wertinek_2018,Newman_2020,Wiedmann_2020}, magnets properties for memory hard-drive \cite{Anders_2020}, and possibly in biological systems \cite{Kolli_2012, Lambert_2013,Scholes_2017,Lambert_2020}. For such applications, since one is usually interested in stationary properties and performances, the knowledge of the steady state of the strongly dissiaptive dynamics is enough.

 Several techniques have been established to gain access to the strong coupling dynamics where usual Markovian Master equations cannot be used straightforwardly. One of them is the polaron transformation \cite{McCutcheon_2010,McCutcheon_2011,Roy_2011,Lee_2012,Gelbwaser_2015,Nazir_2015}, which consists in analyzing the problem in a rotating frame with respect to the coupling Hamiltonian. In the following, however, we will focus on two other approaches which have been recently used to obtain estimates of the steady states in the strong coupling regime. 
The first one uses embedding techniques like reaction coordinate \cite{ Smith_2014,Smith_2016,Strasberg_2016,Strasberg_2018,Correa_2019,Wertinek_2018} and pseudo-mode \cite{Garraway_1997,Pleasance_2017, Teretenkov_2019,Pleasance_2020} to obtain the dynamics of the system of interest and then considers time going to infinity. 
The second one relies on a result sustained by several studies \cite{Bach_2000,Frohlich_2004,Merkli_2007,Mori_2008,Konenberg_2016,Merkli_2020} establishing, under some generic conditions, 
that a system $S$ interacting strongly with a thermal bath $B$ tends together with $B$ to a global thermal state. Greater details on this technique can be found in the recent review \cite{Trushechkin_2021b}.

Both approaches have some strengths and weaknesses. 
The limitations of the first approach relying on embedding techniques typically come from the bath spectral density.
 On the other hand, the results are expected to have a broad range of validity in terms of coupling strength.
 The second approach requires to trace out the bath in the global thermal state, which actually amounts to similar difficulties as computing the exact dynamics in the first place. Thus, one is left with perturbative expansions, with limited range of validity. However, within the range of validity of these expansions, the obtained expressions are expected to provide very good description of the steady states.

   Even though one expects these two approaches to coincide, at least for some regions of parameter, this has not been tested. This is the first aim of this paper.
   Secondly, we will use the strength of each approach to benchmark and define more precisely the range of validity of the other approach.    
   This allows us to introduce and successfully test some simple validity criteria for both approaches, providing some answers to the crucial question of validity range for each approach. 

\section{Perturbative expansion approach}\label{secPE} 
We consider a system $S$, of self Hamiltonian $H_S$, interacting with a thermal bosonic bath $B$ of self Hamiltonian $H_B$ at inverse temperature $\beta := 1/k_BT$ ($T$ being the usual temperature). The interaction is of the form $V=AB$, where $A$ is an observable of $S$, and $B$ is the standard bosonic operator $B=\sum_k g_k (b_k^{\dag} + b_k)$, where $g_k$ is the coupling coefficient between $S$ and the $k^{\rm th}$ mode of the bath (setting $\hbar =1$), with creation and annihilation operators $b_k^{\dag}$ and $b_k$, respectively. 
The starting point of the perturbative expansion approach is the convergence of the system together with the bath towards the global thermal state
\be\label{globalth}
\rho_{SB}^{\rm th} = Z_{SB}^{-1} e^{-\beta H_{SB}},
\ee
where $H_{SB} := H_S+H_B+V$ is the total Hamiltonian generating the dynamics of $SB$, and $Z_{SB}:= {\rm Tr}_{SB}[e^{-\beta H_{SB}}]$ is the partition function. This fundamental result has been widely used for classical and quantum systems, and is supported by several important studies focusing on quantum systems \cite{Bach_2000,Frohlich_2004,Merkli_2007,Mori_2008,Konenberg_2016,Merkli_2020,Trushechkin_2021b} under the important condition $[H_S,V] \ne 0$ (which can be seen as pure dephasing and can be simply solved). 
From \eqref{globalth}, the reduced steady state of $S$ is obtained by tracing out $B$, obtaining the so-called {\it mean force Gibbs state},
\be\label{mainrhosss}
\rho_S^{\rm ss} := {\rm Tr}_B [\rho_{SB}^{\rm th}].
\ee
As usually, tracing out the bath is a very challenging task which can only be done approximately. A common approach is perturbative expansion, whose main steps are presented below (for more details see Appendix \ref{appexp} and \cite{Mori_2008,Trushechkin_2021b, Subasi_2012,Cresser_2021, Purkayastha_2020}).
By ``taking out'' the local contributions in \eqref{mainrhosss} and then expanding up to second order, we obtain,
\begin{widetext}
\bea
\rho_S^{\rm ss} &=& Z_{SB}^{-1}{\rm Tr}_B\left[ e^{-\beta(H_S+H_B)}e^{- {\cal T}\int_0^\beta du \tilde A(u) \tilde B(u)}\right]\nn\\
&\underset{{\rm 2^d~order}}{\simeq}&  Z_{SB}^{-1}e^{-\beta H_S}\left[1 - \int_0^\beta du \tilde A(u){\rm Tr}_B [e^{-\beta H_B} \tilde B(u)] + \int_0^\beta du_1\int_0^{u_1}du_2 \tilde A(u_1)\tilde A(u_2) {\rm Tr}_B [e^{-\beta H_B}\tilde B(u_1)\tilde B(u_2)]\right]\nn\\ 
&=& \frac{Z_SZ_B}{ Z_{SB}}\rho_S^{\rm th}\left[1 + \int_0^\beta du_1\int_0^{u_1}du_2 \tilde A(u_1)\tilde A(u_2) c_B(u_1-u_2)\right], 
\eea
\end{widetext}
where we used in the first line the usual ``splitting" formula \cite{Feynman_1951}, and defined the operators $\tilde X(u):= e^{u(H_S+H_B)}Xe^{-u(H_S+H_B)}$, the bath correlation function $c_B(u_1-u_2):={\rm Tr}_B [\rho_B^{\rm th}\tilde B(u_1)\tilde B(u_2)]={\rm Tr}_B [\rho_B^{\rm th}\tilde B(u_1-u_2)\tilde B]$ taken in the thermal state $\rho_B^{\rm th}:=Z_B^{-1}e^{-\beta H_B}$ with $Z_B:={\rm Tr}_B[e^{-\beta H_B}]$, and the local thermal state of $S$, $\rho_S^{\rm th}=Z_S^{-1}e^{-\beta H_S}$, with the local partition function $Z_S:={\rm Tr}_S[e^{-\beta H_S}]$. We also use the property of stationary baths, namely, ${\rm Tr}_B [\rho_B^{\rm th} \tilde B(u)]=0$. We obtain for the bath correlation function,
\be\label{maincb}
c_B(u)  = \int_0^\infty d\omega J(\omega) \left[ e^{-\omega u}(n_\omega +1) + e^{\omega u} n_\omega\right]
\ee
where $n_\omega^{\rm th} = (e^{\omega\beta}-1)^{-1}$ is the thermal occupation at the frequency $\omega$, and the bath spectral density is defined as
\be
J(\omega) := \sum_k g_k^2\delta(\omega -\omega_k).
\ee
Introducing the eigen-decomposition of the coupling observable $A=\sum_{\nu} A(\nu)$ such that $[A(\nu),H_S]=\nu A(\nu)$, $A^{\dag}(\nu) = A(-\nu)$, and $\tilde A(u)= \sum_{\nu} e^{-\nu u}A(\nu)$, we have, up to the second order,
\bea\label{mainss}
\rho_S^{\rm ss, PE} \underset{{\rm 2^d~order}}{=} \frac{Z_SZ_B}{ Z_{SB}}\rho_S^{\rm th}\left[1+\sum_{\nu,\nu'}A(\nu)A^{\dag}(\nu') g(\nu,\nu') \right], \nn\\
\eea
where 
\be\label{gfct}
g(\nu,\nu'):= \int_0^\beta du_1\int_0^{u_1}du_2 e^{-\nu u_1 +\nu' u_2}c_B(u_1-u_2),
\ee
and the superscript ``PE" stands for ``Perturbative Expansion". 
 Addtionally,
\bea
Z_{SB}&=&{\rm Tr}_{SB}[e^{-\beta(H_S+H_B+V)}]\nn\\
&\underset{{\rm 2^d~order}}{=}& Z_SZ_B\left[1+\sum_{\nu}{\rm Tr}_S[\rho_S^{\rm th}A(\nu)A^{\dag}(\nu)] g(\nu,\nu) \right]. \nn\\
\eea
Expression \eqref{mainss} is Hermitian and is equivalent to the expression obtained in \cite{Cresser_2021} (up to the initial renormalisation term). 
An explicit, analytical expression of the function $g(\nu,\nu')$ in term of usual functions is provided in Appendix \ref{appexp} for under-damped \eqref{JUD} and over-damped \eqref{JOD} bath spectral densities.

\subsection{Conditions of validity}\label{secvalidity}
The above expression \eqref{mainss} is a second order expansion. It provides a good approximation of the exact steady state $\rho_S^{\rm ss}$ Eq.\eqref{mainrhosss}  in the limit of small corrections, when the second order contribution is much larger than higher order contributions. Based on that, one can build validity criteria by requiring that the corrections brought by \eqref{mainss} with respect to the thermal state $\rho_S^{\rm th}=Z_S^{-1}e^{-\beta H_S}$ remain small. Expressed in different mathematical ways, we present in the following several potential validity criteria to be tested later with numerical simulations (section \ref{seccomp}).  
%
%
%
\begin{itemize}
\item A first criterion can be obtained by considering that small corrections should imply that the global partition function $Z_{SB}$ is close the product of the local partition functions $Z_SZ_B$,
\be\label{cr1}
{\rm cr}_1:= \Big|\frac{Z_{SB}}{Z_BZ_S} - 1\Big| \ll1,
 \ee
which turns out to be equivalent to the condition suggested in \cite{Cresser_2021}. 

\item Alternatively, one could consider that the expansion is valid as long as the corrections to the populations (in the eigenbasis of $H_S$) are small, resulting in the following criterion
\be\label{cr2}
{\rm cr}_2:= \frac{\Big|p_n^{\rm ss} - p_n^{\rm th}\Big|}{p_n^{\rm ss}} \ll 1,
\ee
for all energy level $n$ of $S$, where $p_n^{\rm ss}$ stands for the populations of the steady state \eqref{mainss} and $p_n^{\rm th}$ corresponds to the populations of the thermal state $\rho_S^{\rm th}$, reached in the weak coupling limit. Note that a coherence-based criterion would typically be equivalent to the above population-based one.

 \item Additionally, the quantity 
 \be\label{defQ}
 Q:=\int_0^\infty d\omega \frac{J(\omega)}{\omega},
 \ee
  known as the ``re-organisation energy'' \cite{Wu_2010,Ritschel_2011, May_2011}, gives a figure of merit of the coupling energy. Therefore, one can expects the expansion to be valid for $Q  \ll \omega_S$,
where $\omega_S$ stands for the typical energy difference in $H_S$. Thus, we define a third candidate for the validity criterion as 
 \be\label{cr3}
 {\rm cr}_3:= \frac{Q}{\omega_S} \ll1.
 \ee  
Anticipating sections \ref{secRC} and \ref{seccomp}, we can obtain explicit expressions of $Q$ in term of the bath parameters for the bath spectral densities used there. For the under-damped spectral density $J_{UD}(\omega) := \omega\frac{2}{\pi}\frac{\gamma_{UD}\Omega^2\lambda^2}{(\Omega^2-\omega^2)^2+(\gamma_{UD}\Omega\omega)^2}$ (see more details in section \ref{secRC}), we obtained $Q_{UD} = \lambda^2/\Omega$, while for the over-damped spectral density $J_{OD}(\omega) := \alpha\omega\frac{\omega_c^2}{\omega_c^2+\omega^2}$, we have $Q_{OD}=\frac{\pi}{2}\alpha\omega_c$.

 \item Finally, considering a two-level system, focus of the comparison section \ref{seccomp}, we can come up with an additional criteria obtained from a special choice of the system parameters for which the partition function can be easily computed. 
  More precisely, if we take $r_x=r_y=0$ (see next section \ref{secspinboson}), we can diagonalise the total Hamiltonian and compute exactly the partition function, giving a simple but non-trivial expression, $Z_{SB}=e^{\beta Q}Z_SZ_B$.
  Then, in the same spirit as in the first criterion, we can consider that small corrections imply $|Z_{SB}/Z_SZ_B -1 | \ll 1$, which leads to the simple criterion
 \be\label{cr4}
{\rm cr}_4:=  \beta Q \ll 1.
 \ee
 Note that at this stage we only use the setting $r_x=r_y=0$ as a mathematical trick to compute $Z_{SB}$, while in the reminder of the paper we consider $r_x \ne 0$, implying $[H_S,V]\ne 0$, necessary condition for the applicability of  \eqref{globalth}.
Although it might not seem totally justified to approximate the order of magnitude of $Z_{SB}$ for arbitrary $r_x$ and $r_y$ by its value for $r_x=r_y=0$, we will see in the following (section IV), that ${\rm cr}_4$ is always very close to ${\rm cr}_1$, justifying afterwards this approximation. 
 Additionally, the factor $e^{\beta Q}$ is reminiscent of the renormalization factor $e^{\beta Q A^2}$ due to the bath interaction \cite{Cresser_2021,CLL_2021}, so one can conjecture that this criterion could be extended to arbitrary systems in the form ${\rm cr}_4 := \beta Q |A^2| \ll1$.
Finally, this last criterion seems promising because it involves the bath inverse temperature $\beta$. Indeed, the expansion \eqref{mainss} becomes trivially valid when the energy scale set by the bath temperature, $k_BT=\beta^{-1}$, is much larger than the system-bath coupling \cite{Cresser_2021} (infinite temperature limit), suggesting that $\beta$ should play a role in the validity criterion.
  \end{itemize}
 
 
 We will test these criteria in section \ref{seccomp} and see that two of them, ${\rm cr}_1$ and ${\rm cr}_4$, seems to indicate particularly well the validity range of expression \eqref{mainss}.


\subsection{Spin-boson model}\label{secspinboson}
In order to obtain explicit comparison with embedding techniques (reaction coordinate), we choose a specific system, namely the spin-boson model, for being a widely used system, experimentally as well as theoretically. The Hamiltonian of the two-level system is of the form 
 \be
 H_S = \frac{\omega_S}{2}(r_x\sigma_x +r_y\sigma_y + r_z\sigma_z)    = \frac{\omega_s}{2} \vec r.\vec \sigma,
 \ee
 where $\vec r$ is a real unit vector of component $r_x$, $r_y$, and $r_z$ (such that $r_x^2+r_y^2 + r_z^2=1$), and $\vec \sigma$ is the Pauli vector of component the Pauli matrices $\sigma_x$, $\sigma_y$, and $\sigma_z$. Importantly, we will use in the following the notation $r:= r_x+i r_y$.  We consider a typical coupling with the bath, namely $A = \sigma_z$. The eigen-decomposition takes the form $A(u) =  A(\omega_s)e^{-u\omega_s} + A(-\omega_s)e^{u\omega_s} +A(0)$, with
\bea
A(\omega_s) &=& -r |g\ket\bra e|,\nn\\
A(-\omega_s) &=& -r^{*} |e\ket\bra g|,\nn\\
A(0) &=& r_z (|e\ket\bra e| - |g\ket\bra g|) := r_z\Sigma_z,
\eea
where 
\bea
&&|e\ket:= \frac{(1+r_z)|+\ket +r |-\ket}{\sqrt{2(1+r_z)}},\nn\\
&&|g\ket:= \frac{-r^{*}|+\ket +(1+r_z) |-\ket}{\sqrt{2(1+r_z)}},
\eea
are the excited and ground eigenstate of $H_S$, respectively. In the above expression, we used the notation $|+\ket$, $|-\ket$ to denote respectively the excited and ground state of $\sigma_z$.  
%
 Injecting these expressions in \eqref{mainss} with the use of the explicit expression of the function $g(\nu,\nu')$ provided in Appendix \ref{appexp}, we obtain for the reduced steady state of $S$ in the basis $\{|e\ket,|g\ket\}$,
\be\label{mainsseg}
\rho_S^{\rm ss, PE} = \begin{pmatrix} p^{\rm ss}_e & c_{ge}^{\rm ss*} \\
                                                      c_{ge}^{\rm ss} & p_g^{\rm ss} 
                                                      \end{pmatrix},
\ee
with
\bea\label{maincss}
c_{ge}^{\rm ss, PE} &=& \frac{-2r r_z(\beta/\omega_s)[G(\omega_s,\beta)- (1+e^{-\omega_s\beta})Q/\beta]}{(1+e^{-\omega_s\beta})[1+r_z^2\beta Q] + |r|^2\beta^2G(\omega_s,\beta)},\nn\\
\eea
and the population,
\bea\label{mainp1ss}
p_e^{\rm ss, PE} &=& \frac{e^{-\omega_s\beta}(1+r_z^2\beta Q)-|r|^2\beta G'(\omega_s,\beta)}{(1+e^{-\omega_s\beta})[1+r_z^2\beta Q] + |r|^2\beta^2G(\omega_s,\beta)},\nn\\
\eea
%
where $Q$ is the re-organisation energy defined above, 
\be
G(\omega_s,\beta)  : = \int_0^1du e^{-\omega_s\beta u} c_B(u\beta),
 \ee
 and $G'(\omega_s,\beta) :=\frac{\partial}{\partial \omega_s}G(\omega_s,\beta)$.
The explicit expression of $G(\omega_s,\beta)$ and $G'(\omega_s,\beta)$ in term of usual functions is provided in Appendix \ref{appexp} for both under-damped and over-damped spectral densities $J_{UD}(\omega)$ \eqref{JUD} and $J_{OD}(\omega)$ \eqref{JOD}.\\


\section{Reaction coordinate}\label{secRC} 
In the perspective of comparing the perturbative expansion approach with embedding approaches, we briefly review some important features of the reaction coordinate mapping. Introduced in \cite{Garg_1985} and further developed in \cite{Huh_2014, Smith_2014,Smith_2016,Strasberg_2016,Strasberg_2018,Nazir_2018}, the archetypal application of reaction coordinate is for the spin-boson model, although it can applied to other systems \cite{Correa_2019,Wertinek_2018,Strasberg_2018}. 
Thus, considering the two-level system of the previous section, the spin-boson model of Hamiltonian $H_{SB} = \frac{\omega_s}{2}\vec r. \vec \sigma + \sigma_z B + H_B$ can be mapped onto \cite{Smith_2014,Smith_2016}
\bea
H_{SB} &=& H_{SRCE}:= \frac{\omega_s}{2}\vec r. \vec \sigma + \lambda \sigma_z(a^{\dag}+a) + \Omega a^{\dag}a \nn\\
&&+ (a^{\dag} + a)B_E + H_E + (a^{\dag} + a)^2\sum_k\frac{g_k^2}{\omega_k},
\eea
where $a$ and $a^{\dag}$ are the annihilation and creation operators of the collective bosonic mode, called the reaction coordinate (RC), defined as $\lambda A(a^{\dag}+a) = \sum_k g_k(b_k^{\dag} + b_k)$, and the system $E$ is a residual bath (the original bath ``minus'' the collective mode) of self Hamiltonian $H_E$ and coupling to the reaction coordinate through the operator $B_E$. The detailed expressions of the residual bath's modes and parameters are not useful in our problem so we refer interested readers to \cite{Smith_2014, Smith_2016} for further details. 
 
 Importantly, the mapping is exact when the original bath $B$ has an under-damped spectral density,
\be\label{JUD}
J_{UD}(\omega) := \omega\frac{2}{\pi}\frac{\gamma_{UD}\Omega^2\lambda^2}{(\Omega^2-\omega^2)^2+(\gamma_{UD}\Omega\omega)^2},
\ee
where $\lambda$ (frequency), $\Omega$ (frequency) and $\gamma_{UD}$ (dimensionless) characterise respectively the strength of the coupling, the peak of the spectral density, and its width. According to the reaction coordinate mapping, the parameters of the collective mode are given directly by the parameters of the under-damped spectral density \cite{Smith_2014,Smith_2016}: $\lambda$ corresponds to the strength of the coupling between $S$ and the collective mode, and $\Omega$ is its frequency.

It is also possible to find an approximate mapping when the original bath spectral density is over-damped, namely of the form,
\be\label{JOD}
J_{OD}(\omega) =\alpha \omega \frac{ \omega_c^2 }{\omega_c^2 +\omega^2}.
\ee
where $\omega_c$ is sometimes referred to as the cutoff frequency and $\alpha$ is a dimensionless parameter determining the coupling strength. 
The $RC$ coupling and frequency can be expressed in terms of the parameters of $J_{OD}(\omega)$ as \cite{Smith_2014,Smith_2016}
\be\label{mainOmega}
\Omega =  \gamma \omega_c ~~ {\rm and}~~ \lambda = \sqrt{\frac{\pi}{2}\alpha\omega_c\Omega}. 
\ee
Written directly in term of the reaction coordinate parameters, the over-damped spectral density takes the form $J_{OD}(\omega) =\omega\frac{2}{\pi} \frac{\lambda^2 \gamma }{\Omega^2 +\gamma^2\omega^2}$.
While the expression \eqref{JOD} is a function with two parameters $\alpha$ and $\omega_c$, this later expression contains three parameters. The extra parameter $\gamma$ is introduced during the reaction coordinate mapping and is required to be much larger than 1. 
 To understand better the emergence of this free parameter $\gamma$, one should mention that for over-damped spectral density, the mapping is actually obtained from an asymptotic limit of the under-damped spectral density case, as follows \cite{Smith_2014,Smith_2016}. For $\gamma_{UD} \gg1$, we have $J_{UD}(\omega) \simeq J_{OD}(\omega)$ when setting $\alpha = \frac{2\gamma_{UD}\lambda^2}{\pi \Omega^2}$ and $\omega_c = \frac{\Omega}{\gamma_{UD}}$. Then, with these settings, we can use the reaction coordinate mapping used for the under-damped case, and the parameter $\gamma$ appearing in \eqref{mainOmega} is actually $\gamma_{UD}$ which must be much larger than 1 in order to have the approximate identification  $J_{UD}(\omega) \simeq J_{OD}(\omega)$ . As a conclusion, the reaction coordinate mapping is not exact for over-damped spectral density, and only holds under the condition $\Omega \gg \omega_c$ (or $\gamma_{UD} \gg1$).  \\

Now that we have introduced the reaction coordinate mapping for the under-damped and over-damped spectral densities, we can focus on the steady state. For weak coupling between the reaction coordinate $RC$ and the residual bath $E$, which is precisely the situation where the reaction coordinate mapping is useful, one expects from weak dissipation theory that the extended system $SRC$ ($S$ and the reaction coordinate) tends to the thermal state at inverse temperature $\beta$, the inverse temperature of the original and residual bath,
 \be\label{srcth}
\rho_{SRC}^{\rm th} = Z_{SRC}^{-1}e^{-\beta H_{SRC}},
\ee
where 
\be\label{srcH}
H_{SRC} := H_S + \lambda A(a^{\dag}+a) + \Omega a^{\dag}a.
\ee
This conjecture was indeed benchmarked by numerical techniques (hierarchical equation of motions) in \cite{Smith_2014,Smith_2016} and Redfield master equation \cite{Purkayastha_2020} and used in \cite{Strasberg_2016,Newman_2017}. However, when the residual coupling between $RC$ and $E$ is not weak, one expects $\rho_{SRC}^{\rm th}$ to depart from the exact steady state of $SRC$. Thus, $\rho_{SRC}^{\rm th}$ becomes an approximation of the exact steady state. How good is this approximation and when exactly does it start breaking down are the questions which motivated this paper. 
In the following, we will refer to $\rho_{SRC}^{\rm th}$ as the ``{\it reaction coordinate mapping of the steady state}'', or ``{\it mapping of the steady state}'' in short.
From $\rho_{SRC}^{\rm th}$, the reduced steady state of $S$ is given by 
\be\label{mainssET}
\rho_S^{\rm ss, RC} := {\rm Tr}_{RC}[\rho_{SRC}^{\rm th}].
\ee
Again, we stress that since in general $\rho_{SRC}^{\rm th}$ is an approximation of the exact steady state of $SRC$, $\rho_S^{\rm ss, RC} $ is also in general an approximation of $\rho_S^{\rm ss}$ \eqref{mainrhosss}, the exact steady state of $S$.

 The partial trace over the RC mode can be realised numerically or analytically via approximate diagonalization of $H_{SRC}$ (see for instance \cite{Irish_2005, Li_2021}). Note that the plots presented below were indeed realised using numerical diagonalisation using QuTiP (with adequate truncation of the RC mode).
The remainder of the paper is mainly dedicated to the comparison of the predictions of the two approaches, namely comparing \eqref{mainssET} with \eqref{mainsseg}, \eqref{mainp1ss}, and \eqref{maincss}. Before that, we introduce a third approximation of the steady state which will help us in the comparison and is detailed in the next section \ref{secexpRC}. 

\subsection{Perturbative expansion applied to reaction coordinate}\label{secexpRC}
 Beyond our prime objective to confirm that the perturbative expansion approach and the reaction coordinate-based approach coincide, at least for some range of parameters, we also aim at studying the validity range of each approach. In that perspective, when some discrepancies appear between $\rho_S^{\rm ss, RC}$ and $\rho_S^{\rm ss, PE}$, how can we tell that it is because the reaction coordinate mapping of the steady state, $\rho_{SRC}^{\rm th}$, fails to faithfully approximate the exact steady state of $RC$, or that it is because the perturbative expansion stops being valid, or both? How can we separate the two effects? 
 
 We can obtain some insights on these questions by considering a third state, obtained by applying the general perturbative expansion of section \ref{secPE} to $\rho_{SRC}^{\rm th}$ \eqref{srcth}, where the reaction coordinate $RC$ plays the role of the bath $B$. We denote the resulting state by $\rho_S^{\rm ss, PRC}$, where the superscript ``PRC" stands for Perturbative expansion of the RC mapping. 
  Then, from the point of view of the perturbative expansion, $\rho_S^{\rm ss, RC}$ is the ``exact'' (containing all orders) version of $\rho_S^{\rm ss, PRC}$. Consequently, the discrepancy between $\rho_S^{\rm ss, RC}$ and $\rho_S^{\rm ss, PRC}$ provides information on the validity of the  perturbative expansion.
  Conversely, $\rho_S^{\rm ss, PE}$ and $\rho_S^{\rm ss, PRC}$ are both expansions to the same order, of the original problem and of the reaction coordinate mapping, respectively. Then, from the point of view of the reaction coordinate mapping, $\rho_S^{\rm ss, PE}$ is the exact version of $\rho_S^{\rm ss, PRC}$. Thus, by observing the discrepancy between $\rho_S^{\rm ss, PE}$ and $\rho_S^{\rm ss, PRC}$ we can obtain information on the performance of the reaction coordinate mapping.
  We will use in the next section these discrepancies $\rho_S^{\rm ss, RC}$ versus $\rho_S^{\rm ss, PRC}$ and $\rho_S^{\rm ss, PE}$ versus $\rho_S^{\rm ss, PRC}$ to gain precious information on the range of validity of each approach.

 Applying the general perturbative expansion of section \ref{secPE} to $\rho_{SRC}^{\rm th}$ we obtain the same form as \eqref{mainss}, namely,
 \bea\label{ssPRC}
\rho_S^{\rm ss, PRC} \underset{{\rm 2^d ~order}}{=} \frac{Z_SZ_B}{ Z_{SB}}\rho_S^{\rm th}\left[1+\sum_{\nu,\nu'}A(\nu)A^{\dag}(\nu') g(\nu,\nu') \right].\nn\\
\eea
 The functions $g(\nu,\nu')$, $c_B(u)$, $G(\nu,\beta)$, and $G'(\nu,\beta)$, have the same general expression as the one detailed in Appendix \ref{appexp} but using the following spectral density
\be
J(\omega) = \lambda^2\delta(\omega-\Omega).
\ee
Thus, for the steady state populations and coherences, it leads to the same expressions as \eqref{mainp1ss} and \eqref{maincss}, respectively, substituting $G(\nu,\beta)$ and $G'(\nu,\beta)$ by $G^{\rm PRC}(\nu,\beta) := C^{\rm PRC}(-\nu) +e^{-\nu\beta}C^{\rm PRC}(\nu)$, and $G^{\rm PRC '}(\nu,\beta):= -C^{\rm PRC '}(-\nu) -\beta e^{-\nu\beta}C^{\rm PRC}(\nu) + e^{-\nu\beta}C^{\rm PRC '}(\nu)$, with
\be
C^{\rm PRC}(\nu) := \frac{\lambda^2}{\beta}\left(\frac{n_{\Omega}^{\rm th} +1}{\Omega-\nu}-\frac{n_{\Omega}^{\rm th}}{\Omega+\nu}\right)
\ee
and
\be
C^{\rm PRC '}(\nu) := \frac{\lambda^2}{\beta}\left(\frac{n_{\Omega}^{\rm th} +1}{(\Omega-\nu)^2}+\frac{n_{\Omega}^{\rm th}}{(\Omega+\nu)^2}\right).
\ee

\begin{figure}
\centering
(a)\includegraphics[width=6.7cm, height=4.3cm]{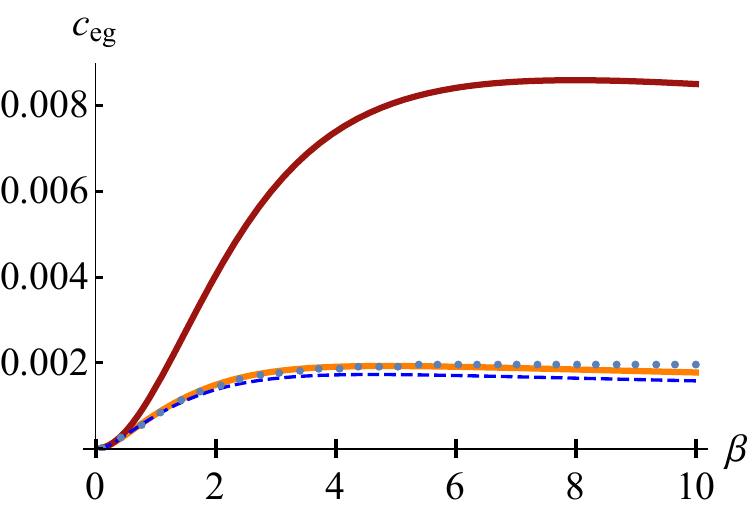}\\
(b)\includegraphics[width=6.7cm, height=4.3cm]{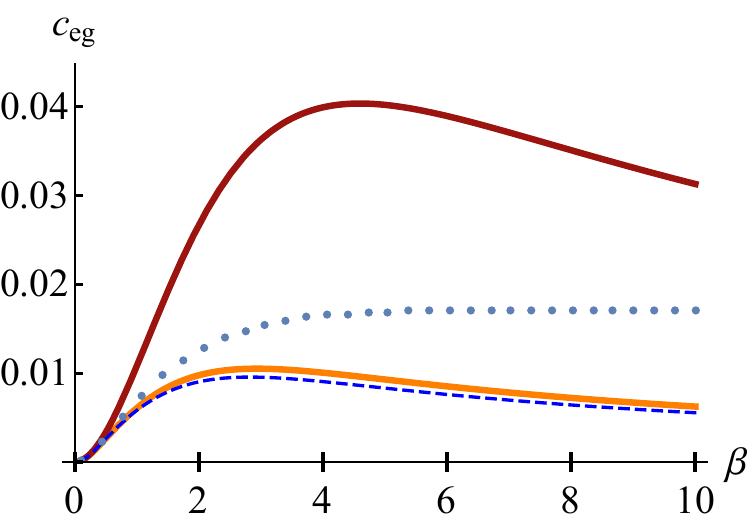}
\caption{Steady state coherences $c_{ge}^{\rm PE, OD}$ \eqref{cOD} (red thick solid line), $c_{ge}^{\rm PE, UD}$ \eqref{cUD} (orange thick solid line), $c_{ge}^{\rm PRC}$ \eqref{cPRC} (blue dashed line), and $c_{ge}^{\rm RC}$ \eqref{cRC} (sparse dotted line), in function of the inverse temperature $\beta$ in unit of $\omega_S^{-1}$, for (a) $\lambda/\omega_S = 0.5$ and (b) $\lambda/\omega_S = 1.5$.  
 The other parameters are given by $\Omega/\omega_S = 10$, $\gamma_{UD}=0.1$, $r_z  = \sqrt{0.75}$, $r = r_x+ir_y = 0.5$.} 
\label{UD_coh}
\end{figure}

  \section{Comparison}\label{seccomp}

In this section, we compare the steady state $\rho_S^{\rm ss, RC}$ given by the reaction coordinate \eqref{mainssET}, the steady state $\rho_S^{\rm ss, PRC}$ \eqref{ssPRC} given by the perturbative expansion of the reaction coordinate, and the steady states $\rho_S^{\rm ss, PE, UD}$ and $\rho_S^{\rm ss, PE, OD}$, given respectively by the perturbative expansion of the original problem \eqref{mainsseg} for under-damped bath spectral density $J_{UD}(\omega)$ \eqref{JUD} and over-damped bath spectral density $J_{OD}(\omega)$ \eqref{JOD}. We denote the coherence and excited population in the eigenbasis $\{|e\ket,|g\ket\}$ of $H_S$ as
\bea
&&p_{e}^{\rm RC}:=\bra e|\rho_S^{\rm ss, RC}|e\ket, \label{pRC}\\
&&c_{ge}^{\rm RC}:=\bra g|\rho_S^{\rm ss, RC}|e\ket, \label{cRC}\\
&&p_{e}^{\rm PRC}:=\bra e|\rho_S^{\rm ss, PRC}|e\ket, \label{pPRC}\\
&&c_{ge}^{\rm PRC}:=\bra g|\rho_S^{\rm ss, PRC}|e\ket, \label{cPRC}\\
&&p_{e}^{\rm PE, UD}:=\bra e|\rho_S^{\rm ss, PE, UD}|e\ket, \label{pUD}\\
&&c_{ge}^{\rm PE, UD}:=\bra g|\rho_S^{\rm ss, PE, UD}|e\ket, \label{cUD}\\
&&p_{e}^{\rm PE, OD}:=\bra e|\rho_S^{\rm ss, PE, OD}|e\ket, \label{pOD}\\
&&c_{ge}^{\rm PE, OD}:=\bra g|\rho_S^{\rm ss, PE, OD}|e\ket. \label{cOD}
\eea

Fig. \ref{UD_coh} presents the plots of the steady state coherences as given by $c_{ge}^{\rm PE, OD}$ (red thick solid line), $c_{ge}^{\rm PE, UD}$ (orange thick solid line), $c_{ge}^{\rm PRC}$ (blue dashed line), and $c_{ge}^{\rm RC}$(sparsely dotted line), in function of the inverse temperature $\beta$ (in unit of $\omega_S^{-1}$). The panel (a) corresponds to a coupling $\lambda/\omega_S =0.5$ and the panel (b) to $\lambda/\omega_S=1.5$. The other parameters are chosen as follows, $\Omega/\omega_S = 10$, $\gamma_{UD}=0.1$ (dimensionless), $r_z= \sqrt{0.75}$, $r =r_x + i r_y = 0.5$, meaning that $r_y = 0$ (note than one would obtain a similar behavior up to a $\pi/2$ phase for the coherence with the alternative choice $r_x = 0$, and $r_y = 0.5$).

One can see a very good agreement between $c_{ge}^{\rm RC}$ (sparse dots) and $c_{ge}^{\rm PRC}$ (blue dashed line) at high temperature (low inverse temperature $\beta$), but this agreement slowly deteriorates beyond $\beta\sim4\omega_S^{-1}$ in panel (a), and beyond $\beta\sim  \omega_S^{-1}$ in panel (b). 
By contrast, the agreement between $c_{ge}^{\rm PE, OD}$ (red solid line) and $c_{ge}^{\rm PRC}$ (blue dashed line) is only good at very high temperature and it deteriorates quickly as the temperature decreases.
Finally, $c_{ge}^{\rm PE, UD}$ (orange solid line) and $c_{ge}^{\rm PRC}$ (blue dashed line) coincide perfectly at high temperature, and only a very small discrepancy appears at small temperatures.   \\

\begin{figure}
\centering
(a)\includegraphics[width=6.3cm, height=3.8cm]{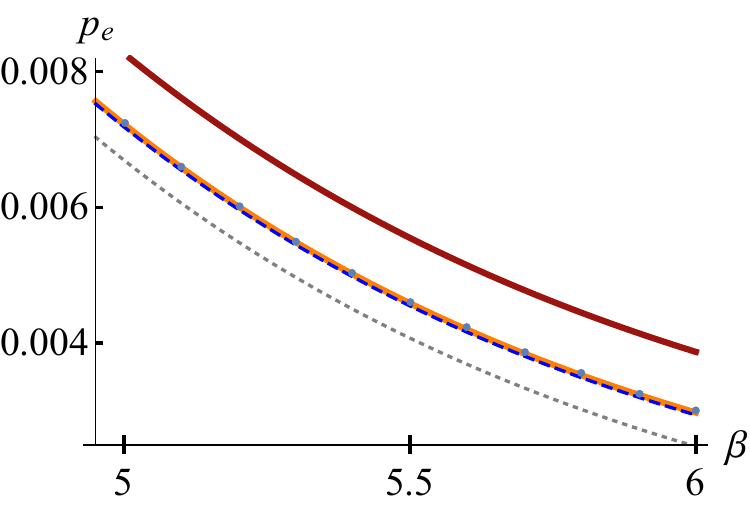}\\
(b)\includegraphics[width=6.3cm, height=3.8cm]{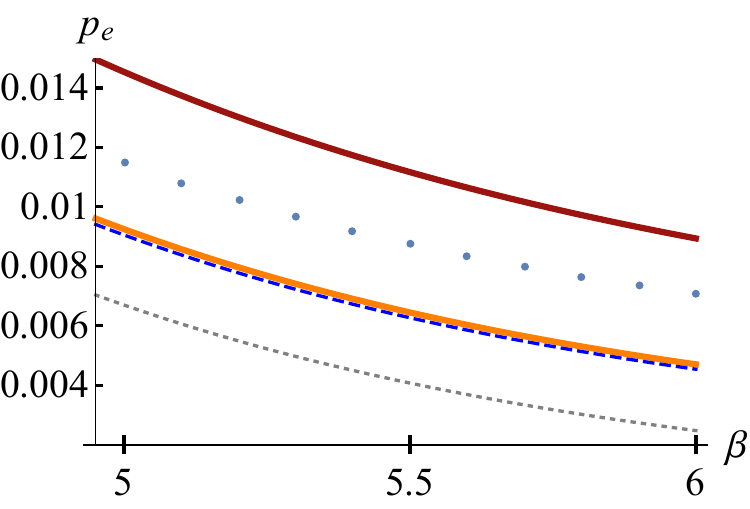}\\
(c)\includegraphics[width=6.3cm, height=3.8cm]{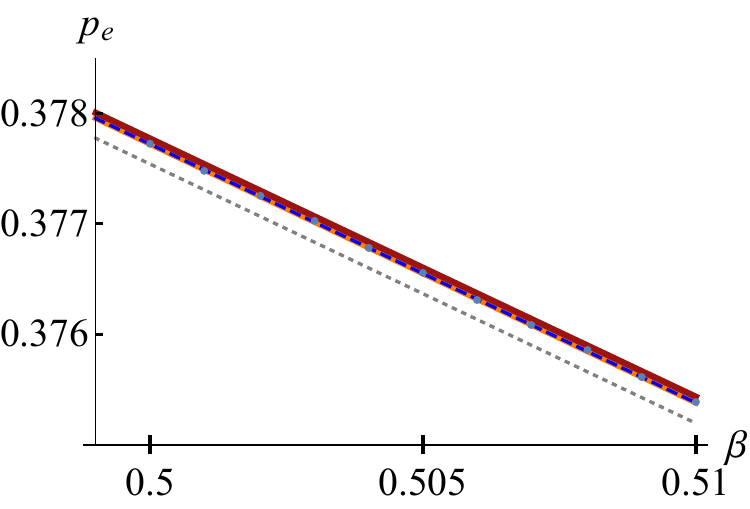}\\
(d)\includegraphics[width=6.3cm, height=3.8cm]{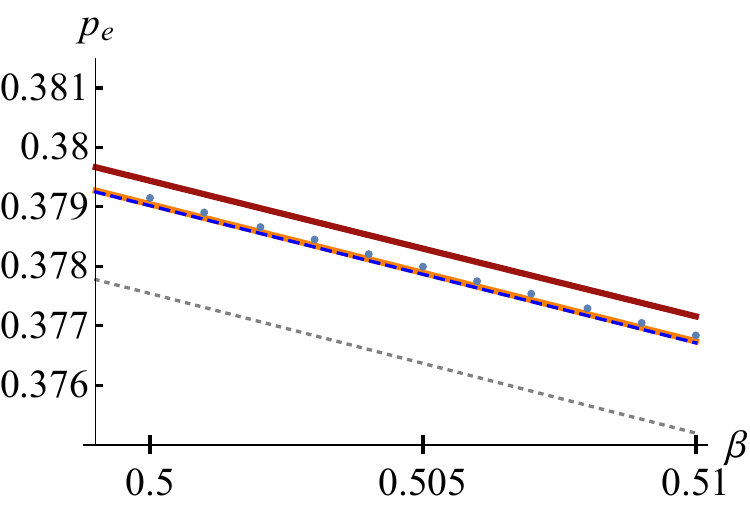}
\caption{Steady state populations in function of the inverse temperature $\beta \in [5;6]$ (in unit of $\omega_S^{-1}$) for panels (a) and (b), and $\beta \in [0.5;0.51]$ for (c) and (d). The coupling strength is $\lambda/\omega_S =0.5$ for panels (a) and (c), and $\lambda/\omega_S =1.5$ for panel (b) and (d). Following the same color convention as in Fig. \ref{UD_coh}, the red thick solid line corresponds to $p_{e}^{\rm PE, OD}$ \eqref{pOD},
the orange thick solid line corresponds to $p_{e}^{\rm PE, OD}$ \eqref{pUD}, the blue dashed line corresponds to $p_{e}^{\rm PRC}$ \eqref{pPRC}, and the sparse dotted line corresponds to $p_{e}^{\rm RC}$ \eqref{pRC}. The dotted grey line represents the thermal excited population $p_e^{\rm th} := \bra e|\rho_S^{\rm th}|e\ket $ at inverse temperature $\beta$. The other parameters are as in Fig \ref{UD_coh}. 
}
\label{UD_pop}
\end{figure}

Fig. \ref{UD_pop} is the counter-part of Fig. \ref{UD_coh} for the excited population. Note that instead of plotting directly the populations for $\omega_S\beta$ from 0 to 10, we zoom in and consider two sections, otherwise all five curves would be indistinguishable: the panels (a) and (b)
represents the steady state excited population as a function of the inverse temperature $\beta$ in the interval $[5;6]$ (in unit of $\omega_S^{-1}$), while panels (c) and (d) correspond to $\beta \in [0.5;0.51]$. Additionally, panels (a) and (c) correspond to a coupling $\lambda/\omega_S =0.5$, while panel (b) and (d) to $\lambda/\omega_S=1.5$. 
 The color convention is the same as in Fig. \ref{UD_coh}, namely, $p_{e}^{\rm PE, OD}$ \eqref{pOD} (red solid line), $p_{e}^{\rm PE, UD}$ \eqref{pUD} (orange solid line), $p_{e}^{\rm PRC}$ \eqref{pPRC} (the blue dashed line), and $p_{e}^{\rm RC}$ (sparsely dotted line) \eqref{pRC}. 
The grey dotted line represents the thermal excited population $p_e^{\rm th} := \bra e|\rho_S^{\rm th}|e\ket $ at inverse temperature $\beta$. The other parameters are chosen as in Fig. \ref{UD_coh}. 

The conclusions are the same as in Fig. \ref{UD_coh}, namely, the agreement between $p_{e}^{\rm RC}$ and $p_{e}^{\rm PRC}$ is very good at high temperature and deteriorates more at low temperature when the system-bath coupling is larger. However, the agreement between $p_{e}^{\rm PE, OD}$ and $p_{e}^{\rm PRC}$ is relatively good only at high temperature, and large discrepancies appear at low temperatures, while $p_{e}^{\rm PE, UD}$ and $p_{e}^{\rm PRC}$ coincide perfectly at both ranges of temperatures, even in strong coupling.\\

\begin{figure}
(a)\includegraphics[width=6.3cm, height=3.9cm]{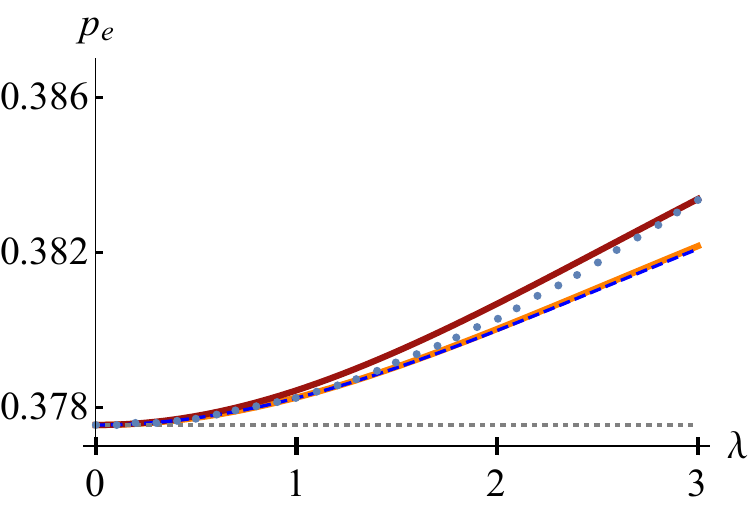}\\
(b)\includegraphics[width=6.3cm, height=3.9cm]{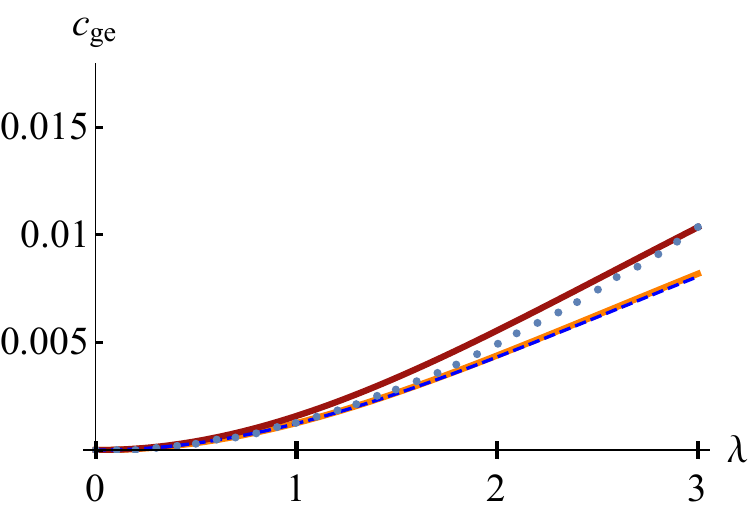}\\
(c)\includegraphics[width=6.3cm, height=3.9cm]{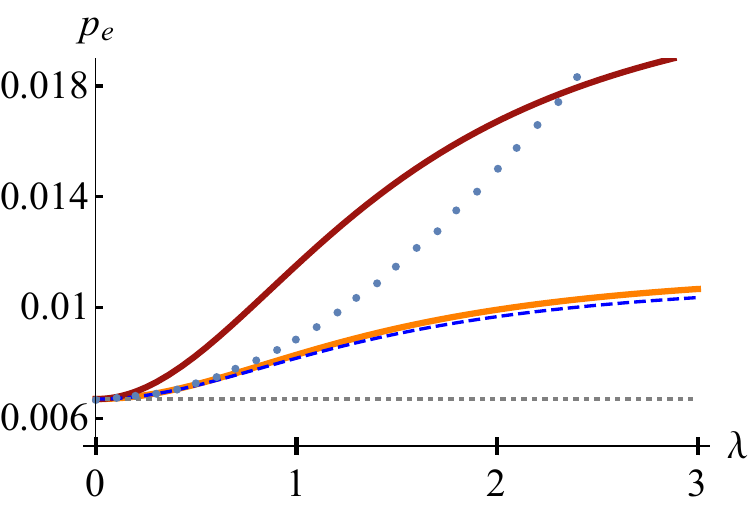}\\
(d)\includegraphics[width=6.3cm, height=3.9cm]{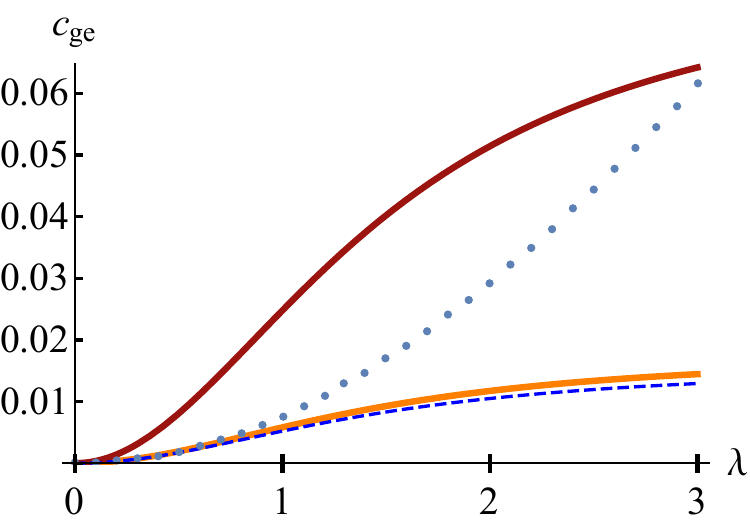}
\caption{Steady state populations (a,c) and coherences (b,d) in function of the coupling strength $\lambda$ (in unit of $\omega_S$) at inverse temperature (a,b) $\omega_S\beta=0.5$ and (c,d) $\omega_S\beta=5$. The color conventions are the same as in Figs. \ref{UD_coh} and \ref{UD_pop}. The remainder of the parameters are chosen as in previous figures, namely $\Omega/\omega_S = 10$, $\gamma_{UD}=0.1$, $r_z =\sqrt{0.75}$, $r=r_x+i r_y = 0.5$.}
\label{UD_popandcoh_la}
\end{figure}

 Finally, Fig. \ref{UD_popandcoh_la} presents the plots of the steady state coherences and populations in function of the coupling strength $\lambda$ (in unit of $\omega_S$) for (a)-(b) $\omega_S\beta = 0.5$ and (c)-(d) $\omega_S\beta = 5$. The color conventions are the same as in Fig. \ref{UD_coh} and Fig. \ref{UD_pop}, as well as the remaining parameters. 
 
\begin{figure}
(a)\includegraphics[width=6.3cm, height=4.2cm]{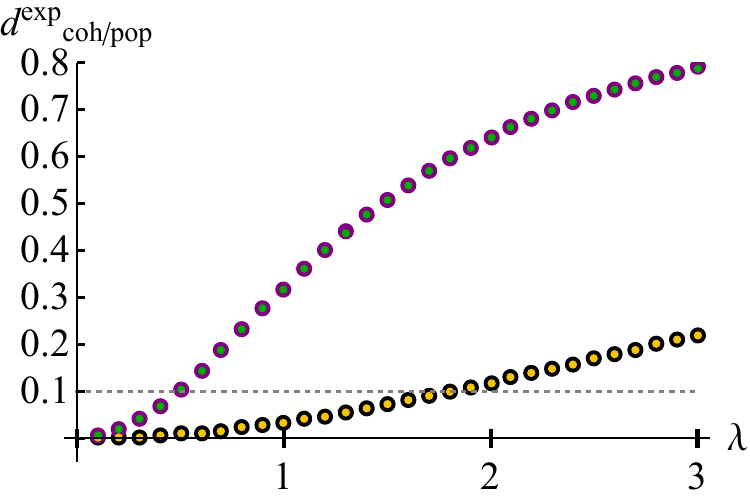}\\
(b)\includegraphics[width=6.3cm, height=4.2cm]{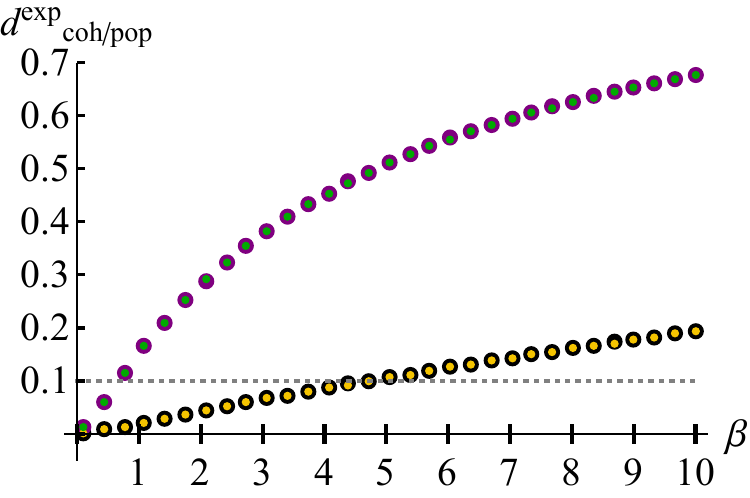}\\
\caption{
Panel (a): the yellow and green dots correspond to $d_{\rm coh}^{\rm exp}$ in function of $\lambda$ (in unit of $\omega_S$) for $\omega_S\beta = 0.5$ and $\omega_S\beta =5$, respectively, and the black and purple large dots (in the background of the yellow and green dots) correspond to $d_{\rm pop}^{\rm exp}$ in function of $\lambda$ for $\omega_S\beta = 0.5$ and $\omega_S\beta =5$, respectively. Panel (b): the yellow and green dots correspond to $d_{\rm coh}^{\rm exp}$ in function of $\beta$ (in unit of $\omega_S^{-1}$) for $\lambda = 0.5\omega_S$ and $\lambda =1.5\omega_S$, respectively, and the black and purple large dots correspond to $d_{\rm pop}^{\rm exp}$ in function of $\beta$ for $\lambda = 0.5\omega_S$ and $\lambda =1.5\omega_S$, respectively.
For both panels, the remainder of the parameters are chosen as in previous figures, namely $\Omega/\omega_S = 10$, $\gamma_{UD}=0.1$, $r_z = \sqrt{0.75}$, $r_x+ir_y = 0.5$}. 
\label{dext}
\end{figure}

 As a first observation, one can see that the plots for the coherences and populations have almost identical shapes, which will be confirmed in Figs. \ref{dext} and \ref{dmap}. Secondly, the agreement between $c_{ge}^{\rm ss, RC}$ (sparse dotted line) and $c_{ge}^{\rm ss, PRC}$ (blue dashed line) (as well as $p_{e}^{\rm ss, RC}$ and $p_{e}^{\rm ss, PRC}$) is excellent below $\lambda \sim 2 \omega_S$ at high temperature $\omega_S\beta =0.5$, and below $\lambda \sim 0.75\omega_S$ at low temperature $\omega_S\beta =5$, while it starts deteriorating beyond these values of coupling strength. 
For $c_{ge}^{\rm PE, OD}$ (red thick line) and $c_{ge}^{\rm PRC}$ (blue dashed line) (as well as $p_{e}^{\rm PE, OD}$ and $p_{e}^{\rm PRC}$), the agreement is good until $\lambda \sim 1$ for $\omega_S\beta=0.5$, but for $\omega_S\beta=5$ the agreement drops quickly beyond $\lambda \sim 0.2 \omega_S$. 
 By contrast, the agreement between $c_{ge}^{\rm PE, UD}$ (orange thick line) and $c_{ge}^{\rm PRC}$ (blue dashed line) (as well as $p_{e}^{\rm PE, UD}$ and $p_{e}^{\rm PRC}$) is almost perfect for both values of $\beta$ and for all $\lambda$. 
 
{\it Conclusion--.} From Fig. \ref{UD_coh} - \ref{UD_popandcoh_la}, we can conclude that the two approaches do coincide on an extended region of parameters for under-damped spectral densities, namely from high to low temperatures for limited coupling $\lambda = 0.5\omega_S$, and even for large coupling $\lambda = 1.5\omega_S$ at high temperature $\omega_S\beta=0.5$. However, for over-damped spectral densities, the two approaches coincide only at high temperatures.

 \subsection{Discrepancies due to the perturbative expansion}
 In order to obtain a more precise and quantitative criterion of ``good" and ``bad" agreement, we introduce the relative discrepancies 
 \bea
 &&d_{\rm coh}^{\rm exp} := (c_{ge}^{\rm RC}-c_{ge}^{\rm PRC})/c_{ge}^{\rm RC},\nn\\
 &&d_{\rm pop}^{\rm exp} := (p_{e}^{\rm RC}-p_{e}^{\rm PRC})/(p_{e}^{\rm RC} - p_{e}^{\rm th}),\nn\\
 \eea
 where one should note that the relative discrepancy related to the population is defined with respect to the deviation from the thermal excited population $p_e^{\rm th} = \bra e|\rho_S^{\rm th}|e\ket $ 
  (at inverse temperature $\beta$). 
As already discussed in section \ref{secexpRC}, these relative discrepancies 
 provide information on the validity of the perturbative expansion. 

Fig. \ref{dext} (a) provides the relative discrepancies associated with Fig. \ref{UD_popandcoh_la}, namely, the yellow and green dots correspond to $d_{\rm coh}^{\rm exp}$ for $\omega_S\beta = 0.5$ and $\omega_S\beta =5$, respectively, while the black and purple large dots (in the background of the yellow and green dots) correspond to $d_{\rm pop}^{\rm exp}$ for $\omega_S\beta = 0.5$ and $\omega_S\beta =5$, respectively. The remainder of the parameters are chosen as in previous figures, namely $\Omega/\omega_S = 10$, $\gamma_{UD}=0.1$, $r_z= \sqrt{0.75}$, $r = r_x+i r_y = 0.5$. Firstly, one can see that the relative discrepancies is exactly the same for coherences and for populations at both temperatures, confirming observations from Fig. \ref{UD_popandcoh_la}. More importantly, adopting the standard 10$\%$ error criterion, these plots testify that the perturbative expansion is valid up to a coupling strength $\lambda \sim 2\omega_S$ at $\omega_S\beta =0.5$, and up to $\lambda \sim 0.5 \omega_S$ at $\omega_S\beta =5$, which also coincides with what we can see from Fig. \ref{UD_popandcoh_la}.

Fig. \ref{dext} (b) provides the relative discrepancies of Figs. \ref{UD_coh} and \ref{UD_pop}. Adopting a similar color convention as panel (a), the yellow and green dots correspond to $d_{\rm coh}^{\rm exp}$ for $\lambda = 0.5\omega_S$ and $\lambda =1.5 \omega_S$, respectively, while the black and purple large dots correspond to $d_{\rm pop}^{\rm exp}$ for $\lambda = 0.5\omega_S$ and $\lambda =1.5\omega_S$, respectively. Thus, as for panel (a), the relative discrepancies are also exactly the same for coherences and populations. \\
{\it Conclusion--.} These plots confirm the observations from Figs. \ref{UD_coh} and \ref{UD_pop}, namely that the perturbative expansion is roughly valid up to $\omega_S\beta \sim 5$ for  $\lambda = 0.5\omega_S$, and up to $\omega_S\beta \sim 1$ for $\lambda = 1.5\omega_S$.

\subsection{Benchmarking the validity criteria for perturbative expansions}\label{seccr}
\begin{figure}[h]
(a)\includegraphics[width=6.7cm, height=4.3cm]{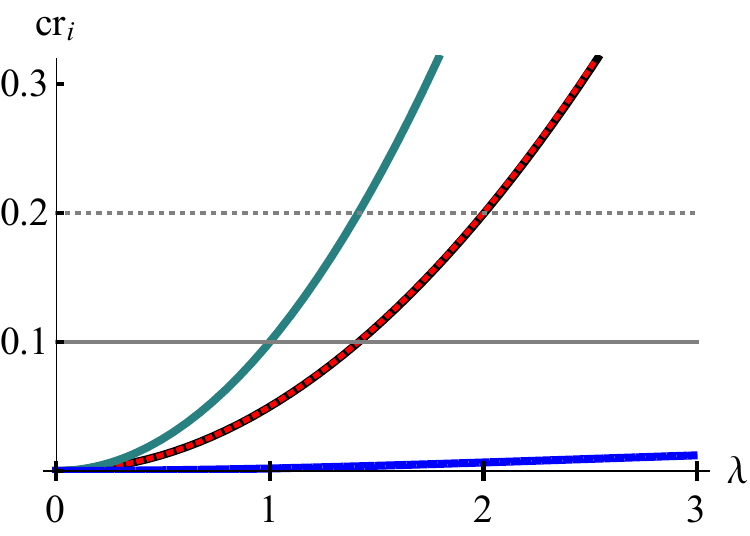}\\
(b)\includegraphics[width=6.7cm, height=4.3cm]{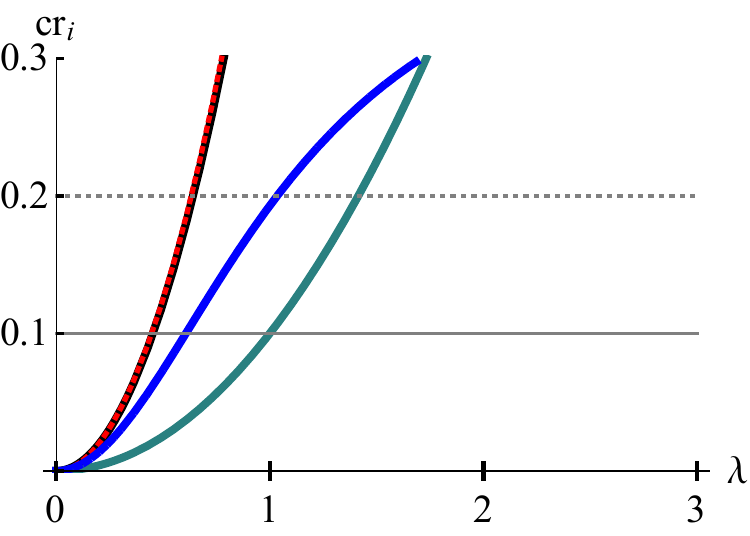}\\
\caption{Plots of the validity criteria introduced in section \ref{secvalidity} in function of the coupling strength $\lambda$ (in unit of $\omega_S$) for (a) $\omega_S\beta=0.5$ and (b) $\omega_S\beta =5$. Criteria ${\rm cr}_1$ and ${\rm cr}_4$ coincide almost exactly and correspond to the black solid line and red dashed line, respectively; ${\rm cr}_2$ corresponds to the blue solid line, and ${\rm cr}_3$ corresponds to the green solid line. The remainder of the parameters are chosen as in previous figures, namely $\Omega/\omega_S = 10$, $\gamma_{UD}=0.1$, $r_z =\sqrt{0.75}$, $r=r_x+i r_y = 0.5$.} 
\label{validity_cr}
\end{figure}

Using the observations from the previous figures, we can benchmark the capacity of the validity criteria introduced in section \ref{secvalidity} to pinpoint the actual range of validity of the perturbative expansion. In Fig. \ref{validity_cr}, we plot the four criteria ${\rm cr}_1$ \eqref{cr1} (black line), ${\rm cr}_2$ \eqref{cr2} (blue line), ${\rm cr}_3$ \eqref{cr3} (green line), ${\rm cr}_4$ \eqref{cr4} (red dashed line), in function of $\lambda$ (in unit of $\omega_S$), for $\omega_S\beta = 0.5$ in panel (a), and $\omega_S\beta=5$ in panel (b). Considering that ${\rm cr}_i \ll 1$ means ${\rm cr}_i \lesssim 0.1$, one can see that only criteria ${\rm cr}_1= \Big|\frac{Z_{SB}}{Z_BZ_S} - 1\Big|$ and ${\rm cr}_4 = \beta Q$, which, interestingly, coincide exactly, indicate a range of validity in agreement with our conclusions from Figs. \ref{UD_popandcoh_la} and \ref{dext}.
More precisely, at inverse temperature $\omega_S\beta =0.5$ ($\omega_S\beta =5$), ${\rm cr}_1$ and ${\rm cr}_4$ indicate a validity of the perturbative expansion up to a coupling strength $\lambda \sim 1.5 \omega_S$ ($\lambda \sim 0.5 \omega_S$), in agreement with the value $\lambda \sim 2 \omega_S$ ($\lambda \sim 0.5 \omega_S$ to $0.75 \omega_S$) from Figs. \ref{UD_popandcoh_la} and \ref{dext}.

\begin{figure}[h]
(a)\includegraphics[width=6.7cm, height=4.3cm]{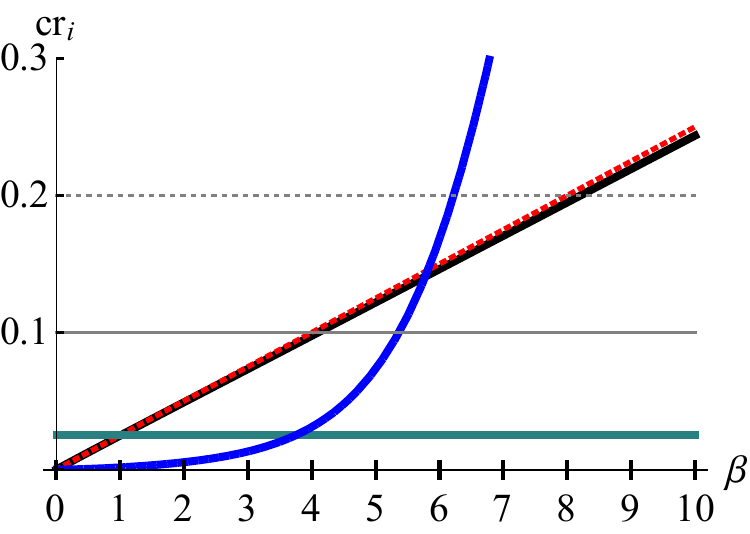}\\
(b)\includegraphics[width=6.7cm, height=4.3cm]{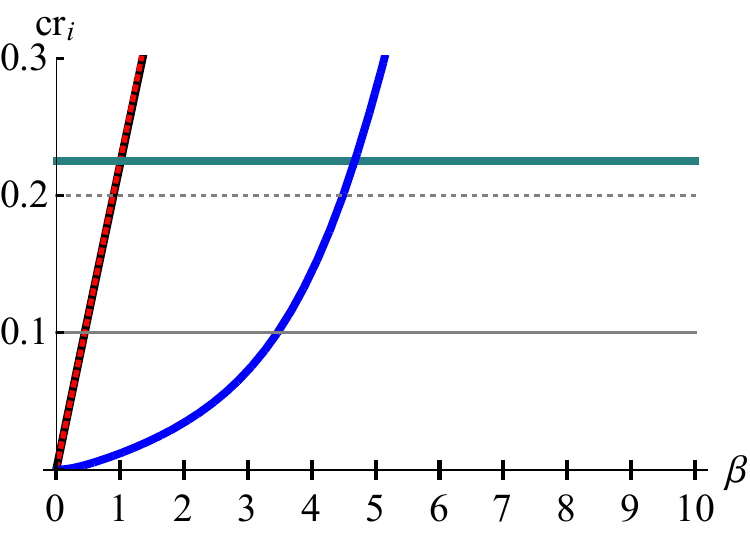}\\
\caption{Plots of the validity criteria introduced in section \ref{secvalidity} in function of $\beta$ (in unit of $\omega_S^{-1}$) for (a) $\lambda=0.5\omega_S$ and (b) $\lambda =5\omega_S$. The color conventions are the same as in Fig. \ref{validity_cr}, namely: ${\rm cr}_1$ (black solid line), ${\rm cr}_4$ (red dashed line), ${\rm cr}_2$ (blue solid line), and ${\rm cr}_3$ (green solid line).
 The remainder of the parameters are chosen as in previous figures, namely $\Omega/\omega_S = 10$, $\gamma_{UD}=0.1$, $r_z =\sqrt{0.75}$, $r=r_x+i r_y = 0.5$.}
\label{cr_vs_b}
\end{figure}

Similarly, in Fig. \ref{cr_vs_b}, we plot the same validity criteria, but now as functions of the inverse temperature, using the same color convention as in the previous figure \ref{validity_cr}. One can see that the conclusions are the same: only ${\rm cr}_1$ and ${\rm cr}_4$, which coincide almost perfectly, indicate a range of validity in agreement with our previous conclusions from Figs. \ref{UD_coh}, \ref{UD_pop} and \ref{dext}. \\

 Thus, we obtain a very simple criterion for the validity of the perturbative expansion incorporating both the coupling strength and the temperature aspects,
 \be\label{finalcr}
 \beta Q \lesssim 0.1
 \ee
This takes the explicit form $\beta \frac{\lambda^2}{\Omega} \lesssim 0.1$ for an under-damped spectral density of parametrization \eqref{JUD}, and $\frac{\pi}{2}\beta \alpha \omega_c \lesssim 0.1$ for an over-damped spectral density of parametrization \eqref{JOD}.

Importantly, we also benchmarked this result for $\Omega \sim \omega_S$ and $\Omega \ll \omega_S$, both in function of $\beta$ and $\lambda$ (see additional plots provided in Appendix \ref{addplots}). In all situations, we confirm that $ \beta Q \lesssim 0.1$ is a surprisingly good validity criterion. In particular, even in regimes where the criterion $Q\ll \omega_S$ is totally misleading, $ \beta Q \lesssim 0.1$ indicates accurately the region where the expansion stops being valid. 

One should note that there is something unexpected in this result. It is comparing the energy scale of the system with the energy scale of the coupling that one sometimes defines weak/strong/ultrastrong coupling: the system's transition energy scale (here $\omega_S$) is the reference. However, our results point at a criterion which is independent of $\omega_S$, namely $ \beta Q \lesssim 0.1$, while the criterion based on $\omega_S$ is wrong most of the time. This could suggest that either the breakdown of the validity of  the perturbative expansion is not strictly related to strong coupling, or the definition of strong coupling might not always be related to the system's energy scale.


 As a side comment, ${\rm cr}_1$ and ${\rm cr}_4$ are exactly equal for $r_x=r_y =0$, by definition. However, their exact agreement, at least for the considered range of parameters (see also plots in Appendix \ref{addplots}), is quite surprising and justifies afterwards the assumption made in section \ref{secvalidity} that the order of magnitude of $Z_{SB}$ does not depend on the spin orientation.

\subsection{Benchmarking the reaction coordinate mapping}
\begin{figure}
(a)\includegraphics[width=6.3cm, height=4.2cm]{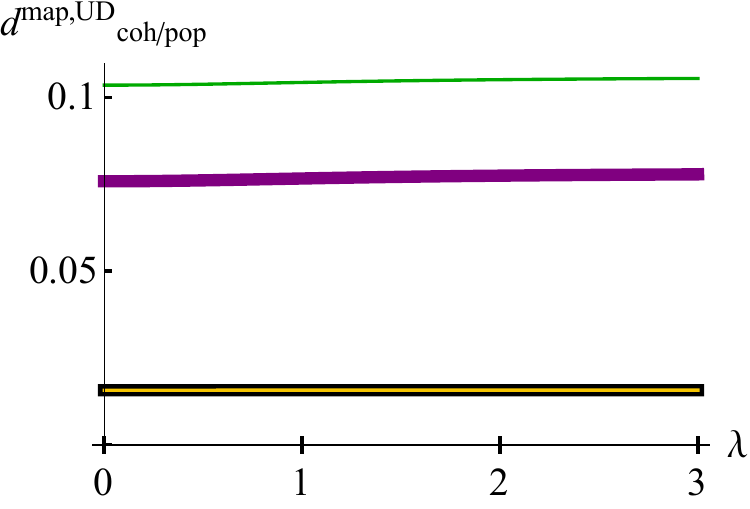}\\
(b)\includegraphics[width=6.3cm, height=4.2cm]{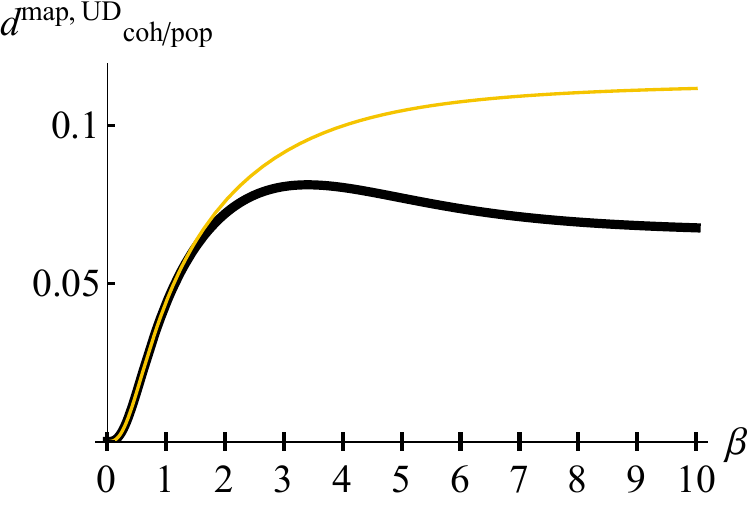}\\
(c)\includegraphics[width=6.3cm, height=4.2cm]{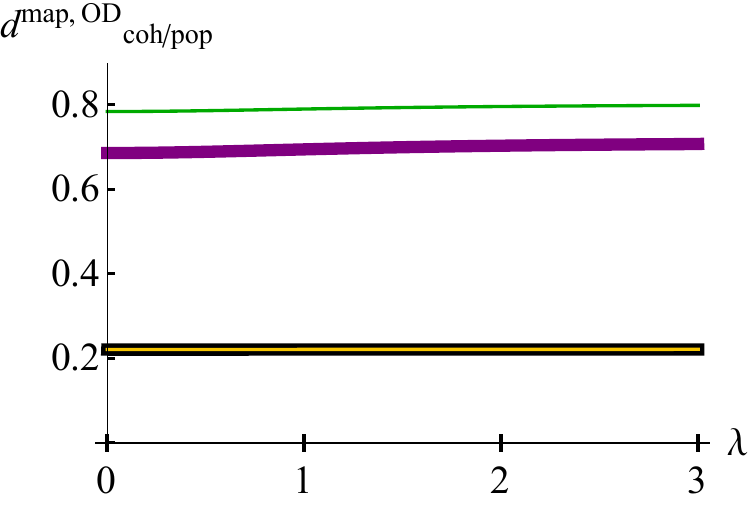}\\
(d)\includegraphics[width=6.3cm, height=4.2cm]{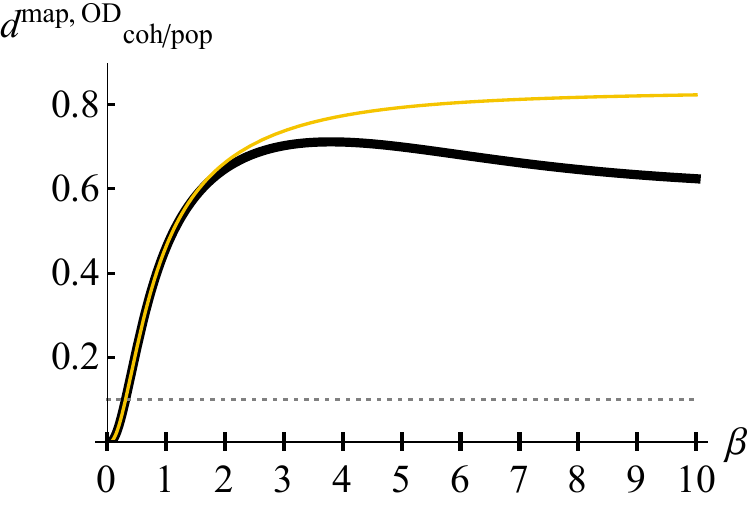}\\
\caption{Plots of the relative discrepancies $d_{\rm pop/coh}^{\rm UD/OD}$ in function of the coupling strength $\lambda$ and $\beta$. Panel (a) (panel (c)): the yellow and green thin solid line represent $d_{\rm coh}^{\rm map, UD}$ ($d_{\rm coh}^{\rm map, OD}$) as a function of $\lambda$ (in unit of $\omega_S$) for $\omega_S\beta = 0.5$ and $\omega_S\beta =5$, respectively, while the black and purple large solid line represent $d_{\rm pop}^{\rm map, UD}$ ($d_{\rm pop}^{\rm map, OD}$) in function of $\lambda$ for $\omega_S\beta = 0.5$ and $\omega_S\beta =5$, respectively.
Panel (b) (panel (d)): the yellow line corresponds to $d_{\rm coh}^{\rm map, UD}$ ($d_{\rm coh}^{\rm map, OD}$) in function of $\beta$ (in unit of $\omega_S^{-1}$) and the black thick line corresponds to $d_{\rm pop}^{\rm map, UD}$ ($d_{\rm pop}^{\rm map, OD}$) also in function of $\beta$, both for arbitrary $\lambda$. 
The other parameters are chosen as in previous figures, namely $\Omega/\omega_S = 10$, $\gamma_{UD}/\omega_S=0.1$, $r_z = \sqrt{0.75}$, $r_x+ir_y = 0.5$.}
\label{dmap}
\end{figure}

In this section we focus on the other aspect of the problem: how well does the reaction coordinate mapping of the steady state \eqref{srcth} approximate the steady state of the original problem, $\rho_S^{\rm ss}$ \eqref{mainrhosss}?
We already saw in Figs. \ref{UD_coh}, \ref{UD_pop}, and \ref{UD_popandcoh_la} that it depends strongly on the bath spectral density as well as on the bath temperature. 
As in the previous section \ref{seccr}, in order to obtain more quantitative information on the performance of the mapping of the steady state, we introduce the following relative discrepancies, 
 \bea
 &&d_{\rm coh}^{\rm map, UD} := (c_{ge}^{\rm PE, UD}-c_{ge}^{\rm PRC})/c_{ge}^{\rm PE, UD},\nn\\ 
 &&d_{\rm pop}^{\rm map, UD} := (p_{e}^{\rm PE, UD}-p_{e}^{\rm PRC})/(p_{e}^{\rm PE, UD} - p_{e}^{\rm th}),\nn\\
  &&d_{\rm coh}^{\rm map, OD} := (c_{ge}^{\rm PE, OD}-c_{ge}^{\rm PRC})/c_{ge}^{\rm PE, OD},\nn\\ 
 &&d_{\rm pop}^{\rm map, OD} := (p_{e}^{\rm PE, OD}-p_{e}^{\rm PRC})/(p_{e}^{\rm PE, OD} - p_{e}^{\rm th}).
 \eea
and plot them in function of $\lambda$ and $\beta$ in Fig. \ref{dmap}.
In Fig. \ref{dmap} (a), the yellow and green thin solid line represent $d_{\rm coh}^{\rm map, UD}$ as a function of $\lambda$ for $\omega_S\beta = 0.5$ and $\omega_S\beta =5$, respectively, while the black and purple large solid line represent $d_{\rm pop}^{\rm map, UD}$ in function of $\lambda$ for $\omega_S\beta = 0.5$ and $\omega_S\beta =5$, respectively. In Fig. \ref{dmap} (c), the same quantities are plotted for the over-damped spectral densities. 
Interestingly, we can see that the relative discrepancies are independent of the coupling strength. In some sense it means that both perturbative expansions $\rho_S^{\rm ss, PRC}$ and $\rho_S^{\rm ss, PE}$ drift away in parallel from their respective exact states $\rho_S^{\rm ss, RC}$ and $\rho_S^{\rm ss}$.
 This is a good indication that the strategy of comparing $\rho_S^{\rm ss, PRC}$ and $\rho_S^{\rm ss, PE}$ does result in getting rid of discrepancies stemming from the perturbative expansion and thus retains only discrepancies stemming from the reaction coordinate mapping, as if we were measuring directly the discrepancies between $\rho_S^{\rm ss, RC}$ and $\rho_S^{\rm ss}$. 
From  Fig. \ref{dmap} (a) we also have the confirmation that the mapping of the steady state performs well for narrow under-damped spectral densities, and fails for over-damped spectral densities, panel (c).

In panel (b), the yellow line corresponds to $d_{\rm coh}^{\rm map, UD}$ in function of $\beta$ (in unit of $\omega_S^{-1}$) and the black thick line corresponds to $d_{\rm pop}^{\rm map, UD}$ also in function of $\beta$ (both for arbitrary $\lambda$ since $d_{\rm pop/coh}^{\rm map, UD}$ is independent of $\lambda$). The same quantities are plotted in panel (d) for over-damped spectral density. 
The main message of these plots is that the reaction coordinate mapping seems to always perform well at high temperatures.

In the following, we briefly explain the reasons behind the failure of the reaction coordinate mapping of the steady state for over-damped spectral densities at arbitrary temperatures, and we also explain its universal success at high temperature.

\subsubsection{Reasons for discrepancies}
Since the reaction coordinate mapping is exact for under-damped bath spectral densities \cite{Smith_2014,Smith_2016,Strasberg_2016}, one might not be surprised that we observed a good performance for such spectral densities. On the other hand, for under-damped spectral densities of increasing width (determined by $\gamma_{UD}$), the reaction coordinate mapping is still exact, but one can verify that the under-damped spectral density becomes indistinguishable from an over-damped spectral density, and the steady state coherences and populations also become indistinguishable from the ones of an over-damped spectral density. 
Thus, how can we have an exact mapping giving a wrong stead state? As already commented in section \ref{secRC},
the reason for this apparent contradiction is that the reaction coordinate mapping of under-damped spectral densities is exact {\it for the dynamics}, and the steady state $\rho_{SRC}^{\rm th}$  \eqref{srcth} is only an approximation of the actual steady state of the dynamics \cite{Smith_2014,Smith_2016,Strasberg_2016,Newman_2017, Purkayastha_2020}. Thus, as already stressed above, the observed failure of the reaction coordinate mapping of the steady state is not a failure of the reaction coordinate mapping per se, but is a break down of the approximation consisting in equating the exact steady state of $SRC$ by $\rho_{SRC}^{\rm th}$ \eqref{srcth}. This breakdown can be understood from three related point of view.
\begin{itemize}
\item  Although the mapping is exact for under-damped spectral densities of arbitrary width, the strength of the coupling between the residual bath and the reaction coordinate grows as $\gamma_{UD}$. Therefore, for increasing spectral widths, one should expect $\rho_{SRC}^{\rm th}$ to depart from the exact steady state. In particular, one expects a steady state of the form $\rho_{SRC}^{\rm ss}  ={\rm Tr}_E[\rho_{SRCE}^{\rm th}] \ne \rho_{SRC}^{\rm th}$. More precisely, one can show \cite{Strasberg_2016} that ${\rm Tr}_{RC}[\rho_{SRC}^{\rm th}]$ is equal to ${\rm Tr}_{B}[\rho_{SB}^{\rm th}]$ at lowest order in the residual coupling between $RC$ and $E$. Thus, for increasing $\gamma_{UD}$, and therefore increasing residual coupling, discrepancies between $\rho_{SRC}^{\rm ss}$ and $\rho_{SRC}^{\rm th} $ as well as $\rho_{S}^{\rm ss}$ and $\rho_{S}^{\rm ss, RC} $ increase.  

\item Alternatively, this can be seen directly from the expressions we obtained for the general perturbative expansion \eqref{mainss}. The steady state \eqref{mainss} depends on the function $g(\nu,\nu')$, which is entirely determined by the bath correlation function $c_B(u)$ \eqref{maincb}, which is itself ultimately determined by the bath spectral density $J(\omega)$. Thus, when approximating the steady state $\rho_S^{\rm ss} ={\rm Tr}_B[\rho_{SB}^{\rm th}]$ \eqref{mainrhosss} by ${\rm Tr}_{RC}[\rho_{SRC}^{\rm th}]$ \eqref{mainssET}, 
 we are ultimately approximating the original bath spectral density by a single mode, represented by the reaction coordinate. In other words, we are approximating the original bath spectral density $J(\omega)$ by $\lambda^2\delta(\omega - \Omega)$. This approximation is reasonable if $J(\omega)$ is a narrow spectral density centered in $\Omega$, but is not justified for a broad spectral density. Thus, one expects that ${\rm Tr}_{RC}[\rho_{SRC}^{\rm th}]$ becomes increasingly distant from $\rho_S^{\rm ss} ={\rm Tr}_B[\rho_{SB}^{\rm th}]$ \eqref{mainrhosss} as the spectral width increases.

%
 
\item  Final viewpoint, strongly related to the first one. One can show that the steady state $\rho_{SRC}^{\rm th}$ \eqref{srcth} is actually the steady state of the master equation derived in the Supplementary Material of \cite{Smith_2014} when applying the secular approximation. However, the secular approximation is valid when ${\rm max}|\omega_{SRC}-\omega_{SRC}'|^{-1} \ll \tau_D$, where $\tau_D$ denotes the dissipation timescale induced by the action of the bath and $\omega_{SRC}$ denotes the Bohr frequencies of the extended system $SRC$.
A rough analysis show that $\tau_D \sim (\pi\gamma_{UD}\omega_S)^{-1} $, so that one expects the secular approximation to become unjustified for growing $\gamma_{UD}$, and thus a steady state increasingly distinct from $\rho_{SRC}^{\rm th}$. This question has been analyzed in great details in \cite{Correa_2019}, and one of the conclusion actually limits the strength of this last argument: the authors show that the reaction coordinate mapping of the steady state might actually be valid way beyond the supposed validity of the secular approximation. 
\end{itemize}

As a rule of thumb, from observations coming from the above plots and additional plots (not shown), one can consider that the reaction coordinate steady state performs well, meaning $d_{\rm pop/coh}^{\rm map, UD} \leq 0.1$, as long as the spectral width $\gamma_{UD}$ is smaller than $ \sim3/\Omega\beta$. 


\subsubsection{Universal faithful mapping of the steady state at high temperature}
\begin{figure}
\includegraphics[width=7.3cm, height=4cm]{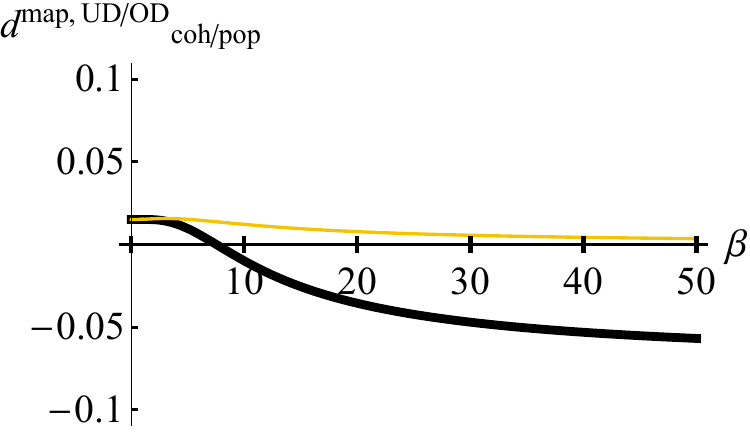}\\
\caption{Plots of $d_{\rm coh}^{\rm map, UD/OD}$ (yellow thin line) and $d_{\rm pop}^{\rm map, UD/OD}$ (black thick line) in function of the inverse temperature $\beta$ (in unit of $\omega_S$) for $\Omega/\omega_S = 1/\beta$, $\gamma_{UD} = 20$ (any value larger than 20 gives the same plot),  $r_z= \sqrt{0.75}$, $r=r_x+ir_y = 0.5$. The plots of $d_{\rm coh}^{\rm map, OD}$ and $d_{\rm coh}^{\rm map, UD}$ ($d_{\rm pop}^{\rm map, OD}$ and $d_{\rm pop}^{\rm map, UD}$) are indistinguishable. Additionally, we chose $\lambda =1.5\omega_S$, but the plots are actually independent of $\lambda$ as shown in Fig. \ref{dmap}.
 }
\label{high_temp_cr}
\end{figure}

Contrasting with the breakdown of the reaction coordinate mapping of the steady state for broad bath spectral densities at arbitrary temperatures, the mapping seems to be always faithful at high temperatures (see Fig. \ref{dmap} (d)). This can be seen as follows.
Assuming that the bath spectral density $J(\omega)$ vanishes for $\omega \geq 2/\beta$, the correlation function can be approximated by
\bea
c_B(u) &:=& {\rm Tr}_B \rho_B^{\rm th} B(u) B\nn\\
 &=& \int_0^\infty d\omega J(\omega) \left[ e^{-\omega u}(n_\omega +1) + e^{\omega u} n_\omega\right]\nn\\
&\sim& \int_0^\infty d\omega J(\omega) \frac{2}{\omega\beta} = \frac{2Q}{\beta},
\eea
where we use the approximation $e^{-\omega u}(n_\omega +1) + e^{\omega u} n_\omega \sim \frac{2}{\omega\beta}$ valid for $\omega\beta\leq 2$ (reminding that the variable $u$ belongs to $[0;\beta]$). 
Then, applying this result to the bath spectral densities we have been considering, $J_{OD}(\omega)$ and $J_{UD}(\omega)$, both vanishing for $\omega \gg \Omega$, one expects to have $c_B(u) \sim\frac{2Q}{\beta}$ for both spectral densities as soon as $\Omega \ll 2/\beta$. Additionally, considering the effective spectral density representing the reaction coordinate $J_{RC}(\omega) = \delta(\omega-\Omega)$, we have $c_B(u) \sim\frac{2Q}{\beta}$ when $\Omega \leq 2/\beta$. 
Thus, for $\Omega \ll 2/\beta$, $G(\omega_s, \beta)$, and therefore $G'(\omega_s,\beta)$, become independent of the form of the bath spectral density, retaining only a dependence on $Q$. 
Then, for $\Omega \ll 2/\beta$, we should have the same steady state for any bath spectral densities of same re-organization energy $Q$. This is what we observed in Fig. \ref{dmap} (d), and confirmed in Fig. \ref{high_temp_cr}, as soon as
\be\label{hightempcr}
\Omega\beta\leq1.
\ee
This hols for arbitrary spectral width $\gamma_{UD}$, and might also hold for arbitrary coupling strength (again, according to our conclusions from Fig. \ref{dmap}).

\section{Conclusion}
We compare the perturbative expansion of the mean force Gibbs state with the approximate steady state \eqref{srcth} from the reaction coordinate mapping.
We reach our first objective by showing the agreement of these two approaches, for some ranges of parameters and focusing on the spin-boson model, see Figs. \ref{UD_coh} - \ref{UD_popandcoh_la}.

In a second time, we focus on the crucial task of exploring and understanding their respective range of validity.
To achieve that, we use one approach to benchmark the other. We establish and test successfully a validity criterion \eqref{finalcr} for the perturbative expansion depending only on the inverse bath temperature $\beta$ and on the reorganization energy $Q$ \eqref{defQ}. 
 
 Regarding the reaction coordinate mapping and its approximate steady state \eqref{mainssET}, we quantify its performance and derived a validity criterion \eqref{hightempcr} involving only the inverse bath temperature $\beta$ and the reaction coordinate frequency $\Omega$, holding for arbitrary spectral width $\gamma_{UD}$ and arbitrary coupling strength. This criterion relies on analytical arguments which were confirmed numerically.
 
  Thanks to these validity criteria, one has in hand practical tools to assess the validity range of these two techniques.
 Although these validity criteria were numerically tested for the spin-boson model, they can be extended to arbitrary systems, and the curious fact that they do not involve the system's energy scale  
  might suggest that they do work for other systems. It would be interesting to test that.
 
 Additionally, it would also be instructive and useful to extend this comparative analysis to other techniques like pseudo-mode \cite{Garraway_1997,Pleasance_2017, Teretenkov_2019,Pleasance_2020}, 
as well as to the ultrastrong coupling regime \cite{Cresser_2021, Trushechkin_2021, CLL_2021}.
\vspace{1cm}

   

\acknowledgements
I am grateful for on going discussions with Ilya Sinayskiy, Graeme Pleasance, and Francesco Petruccione.
I also would like to thank Patrice Camati for a crash course on QuTiP, as well as all QuTiP contributors for setting up and developing such a useful tool. 
The author acknowledges the support of the French Agence Nationale de la Recherche (ANR), under grant ANR-20-ERC9-0010 (project QSTEAM).\\

\appendix
\section{Expression of the function $g(\nu,\nu')$}\label{appexp}

The function $g(\nu,\nu')$ introduced in the main text is defined by $g(\nu,\nu') := \int_0^\beta du_1\int_0^{u_1}du_2 e^{-\nu u_1 +\nu' u_2}c_B(u_1-u_2)$. Re-writing its expression and introducing the variable $v_2 = u_1-u_2$, we obtain
\bea
g(\nu,\nu') &=& \int_0^\beta du_1\int_0^{u_1}du_2 e^{-\nu u_1 +\nu' u_2}c_B(u_1-u_2)\nn\\
&=&\int_0^{\beta}du_1\int_0^{u_1}dv_2 e^{-\nu u_1} e^{\nu'(u_1-v_2)}c_B(v_2)\nn\\
&=&\int_0^{\beta}du_1\int_0^{u_1}dv_2 e^{(\nu'-\nu) u_1} e^{-\nu' v_2}c_B(v_2)\nn\\
&=&\int_0^{\beta}dv_2\int_{v_2}^{\beta}du_1 e^{(\nu'-\nu) u_1} e^{-\nu' v_2}c_B(v_2)\nn\\
&=&\int_0^{\beta}dv\frac{e^{(\nu'-\nu) \beta} - e^{(\nu'-\nu) v} }{\nu'-\nu} e^{-\nu' v}c_B(v)\nn\\
&=&\beta \int_0^{1}dv\frac{e^{(\nu'-\nu) \beta} - e^{(\nu'-\nu) v\beta} }{\nu'-\nu} e^{-\nu' v\beta}c_B(v\beta)\nn\\
&=&\frac{\beta}{\nu'-\nu} \int_0^{1}dv[e^{(\nu'-\nu) \beta} e^{-\nu' v\beta}- e^{-\nu v\beta} ] c_B(v\beta)\nn\\
&=&\frac{\beta}{\nu'-\nu} \Bigg[e^{(\nu'-\nu) \beta} \int_0^{1}due^{-\nu' \beta u}c_B(u\beta)\nn\\
&&\hspace{1.4cm}- \int_0^{1}due^{-\nu \beta u}c_B(u\beta) \Bigg]. \label{A6}
\eea
We are then led to compute 
\be
G(\nu,\beta):=\int_0^1du e^{-\nu\beta u} c_B(u\beta).
\ee
In order to obtain analytical expressions for the under-damped and over-damped bath spectral density, it will be convenient to decompose $G(\nu,\beta)$ in the following way,
\be
G(\nu,\beta) = C(-\nu) + e^{-\beta \nu} C(\nu),
\ee
where
\bea
C(\nu) &:=& \int_0^\infty d\omega J(\omega) \left[ \frac{n_\omega +1}{\beta(\omega-\nu)} -\frac{n_\omega}{\beta(\omega+\nu)} \right]\\
&=& \int_0^\infty d\omega J(\omega) \frac{\nu{\rm coth}(\omega\beta/2) + \omega}{\beta(\omega^2-\nu^2)}.
\eea
Then, from \eqref{A6}, we obtain

\bea
g(\nu,\nu') = \frac{\beta}{\nu'-\nu} \left[e^{(\nu'-\nu)\beta}G(\nu',\beta) - G(\nu,\beta)\right],
\eea
for $\nu\ne\nu'$, and for $\nu'=\nu$,
\be
g(\nu) :=g(\nu,\nu) =\beta\left[ \beta G(\nu,\beta) + G'(\nu,\beta)\right] 
\ee
where $G'(\nu,\beta):= \frac{\partial }{\partial \nu}G(\nu,\beta)= -C'(-\nu) -\beta e^{-\nu\beta}C(\nu) + e^{-\nu\beta}C'(\nu)$ and $C'(\nu)$ is the partial derivative with respect to $\nu$, 
\bea
C'(\nu) &:=& \frac{ \partial C(\nu)}{\partial \nu}\nn\\
 &=& \int_0^\infty d\omega J(\omega) \left[ \frac{n_\omega +1}{\beta(\omega-\nu)^2} +\frac{n_\omega}{\beta(\omega+\nu)^2} \right].\nn\\
\eea
Alternatively, in term of the function $C(\nu)$, we have $g(\nu)= \beta\left[\beta C(-\nu) - C'(-\nu)+e^{-\beta\nu}C'(\nu)\right] $.

\subsection{Exact expression of $C(\omega_S)$ and $C'(\omega_S)$ }
\subsubsection{ Over-damped (Lorentz-Drude) spectral density}\label{appOD}
For $J_{OD}(\omega) = \alpha \frac{\omega \omega_c^2}{\omega^2+\omega_c^2}$, 
 we have
\bea
C_{OD}(\nu) &=& \int_0^\infty d\omega J(\omega) \frac{\nu{\rm coth}(\omega\beta/2) + \omega}{\beta(\omega^2-\nu^2)}\nn\\
&=&\frac{\alpha \omega_c^2}{\beta}\int_0^\infty d\omega \frac{\omega}{\omega^2+\omega_c^2}\frac{\nu \coth{\omega\beta/2} +\omega}{\omega^2-\nu^2}\nn\\
&=&\frac{\alpha \omega_c^2}{\beta}\Bigg[\int_0^\infty d\omega \frac{\omega}{\omega^2+\omega_c^2}\frac{\omega}{\omega^2-\nu^2}\nn\\
&& + \frac{2\nu}{\beta} \sum_{n=-\infty}^{+\infty}\int_0^\infty d\omega \frac{\omega}{\omega_c^2+\omega^2}\frac{1 }{\omega^2-\nu^2}\frac{\omega}{\omega^2+\nu_n^2}\Bigg]\nn\\
\eea
with $\nu_n = 2\pi n/\beta$, called the Matsubara frequencies \cite{FrancescoBook}. For the first term we have
\bea
&&\int_0^\infty d\omega \frac{\omega}{\omega^2+\omega_c^2}\frac{\omega}{\omega^2-\nu^2} \nn\\
&&= \frac{1}{\omega_c^2+\nu^2}\int_0^\infty d\omega\left(\frac{\omega_c^2}{\omega^2 + \omega_c^2} + \frac{\nu^2}{\omega^2-\nu^2}\right)\nn\\
&&= \frac{\omega_c^2}{\omega_c^2+\nu^2}\frac{\pi}{2\omega_c}.
\eea
Generalising that to situations where $\omega_c$ is a complex number (which will be useful for under-damped spectral densities, see in the following), we have 
\bea\label{cpxint}
\int_0^\infty d\omega \frac{\omega}{\omega^2+\omega_c^2}\frac{\omega}{\omega^2-\nu^2} = \begin{cases} \frac{\omega_c^2}{\omega_c^2+\nu^2}\frac{\pi}{2\omega_c} & \text{if  $\Re\omega_c >0$}, \\
 -\frac{\omega_c^2}{\omega_c^2+\nu^2}\frac{\pi}{2\omega_c} & \text{if  $\Re \omega_c <0$}.
 \end{cases}\nn\\
\eea
For the second term, we have 
\begin{widetext}
\bea
&&\sum_{n=-\infty}^{\infty}\int_0^\infty d\omega \frac{\omega}{\omega_c^2+\omega^2}\frac{1 }{\omega^2-\nu^2}\frac{\omega}{\omega^2+\nu_n^2} \nn\\&&= \sum_{n=-\infty}^{\infty}\frac{1}{\nu^2 +\nu_n^2}\int_0^{\infty}d\omega \left[\frac{\nu_n^2}{\omega_c^2-\nu_n^2}\left(\frac{1}{\omega^2+\nu_n^2} -\frac{1}{\omega^2+\omega_c^2}\right) +\frac{\nu^2}{\nu^2+\omega_c^2}\left(\frac{1}{\omega^2-\nu^2}-\frac{1}{\omega^2+\omega_c^2}\right)\right]\nn\\
&&= \sum_{n=-\infty}^{\infty}\frac{1}{\nu^2 +\nu_n^2} \left[\frac{\nu_n^2}{\omega_c^2-\nu_n^2}\left(\frac{\pi}{2|\nu_n|} -\frac{\pi}{2(\pm\omega_c)}\right) +\frac{\nu^2}{\nu^2+\omega_c^2}\left(0-\frac{\pi}{2(\pm\omega_c)}\right)\right]\nn\\
&&= \frac{\pi}{2}\sum_{n=-\infty}^{\infty}\frac{1}{\nu^2 +\nu_n^2}\frac{1}{\pm\omega_c} \left[\frac{|\nu_n|}{\pm\omega_c+|\nu_n|} - \frac{\nu^2}{\nu^2+\omega_c^2}\right]\nn\\
&&= \frac{\pi}{\pm2\omega_c}\left[\sum_{n=-\infty}^{\infty}\frac{1}{\nu^2 +\nu_n^2}\frac{|\nu_n|}{\pm\omega_c+|\nu_n|} - \frac{\nu^2}{\nu^2+\omega_c^2}\sum_{n=-\infty}^{\infty}\frac{1}{\nu^2 +\nu_n^2}\right]\nn\\
&&= \frac{\pi}{\pm2\omega_c}\left[2\sum_{n=1}^{\infty}\frac{1}{\nu^2 +\nu_n^2}\frac{\nu_n}{\pm\omega_c+\nu_n} -\frac{1}{\nu^2+\omega_c^2}- \frac{2\nu^2}{\nu^2+\omega_c^2}\sum_{n=1}^{\infty}\frac{1}{\nu^2 +\nu_n^2}\right]\nn\\
&&= -\frac{\pi}{\pm2\omega_c}\frac{1}{\nu^2+\omega_c^2}+\frac{\pi}{\pm2\omega_c}\left[2\frac{\beta^2}{4\pi^2}\sum_{n=1}^{\infty}\frac{1}{n^2+\frac{ \nu^2\beta^2}{4\pi^2}}\frac{n}{n \pm\frac{\omega_c\beta}{2\pi}} - \frac{2\nu^2}{\nu^2+\omega_c^2}\frac{\beta^2}{4\pi^2}\sum_{n=1}^{\infty}\frac{1}{n^2+\frac{ \nu^2\beta^2}{4\pi^2}}\right]\nn\\
&&= -\frac{\pi}{\pm2\omega_c}\frac{1}{\nu^2+\omega_c^2}+\frac{\pi}{\pm\omega_c}\frac{\beta^2}{4\pi^2}\left[\frac{2\pi^2}{\beta^2}\frac{1}{\nu^2+\omega_c^2}F(\nu\beta,\pm\omega_c\beta) - \frac{\nu^2}{\nu^2+\omega_c^2}\frac{\frac{\nu\beta}{2} \coth\frac{\nu\beta}{2} -1}{\frac{ \nu^2\beta^2}{2\pi^2}}\right]\nn\\
&&= -\frac{\pi}{\pm2\omega_c}\frac{1}{\nu^2+\omega_c^2}+\frac{\pi}{\pm2\omega_c}\frac{1}{\nu^2+\omega_c^2}\left[F(\nu\beta,\pm\omega_c\beta) - \frac{\nu\beta}{2} \coth\frac{\nu\beta}{2} +1\right]\nn\\
&&= \frac{\pi}{\pm2\omega_c}\frac{1}{\nu^2+\omega_c^2}\left[F(\nu\beta,\pm\omega_c\beta) - \frac{\nu\beta}{2} \coth\frac{\nu\beta}{2} \right],\nn\\
\eea
where $\pm\omega_c$ stands for the possibility of $\omega_c$ being complex, where in such case one has to choose the sign corresponding to $\Re (\pm\omega_c)>0$, and 
\be
F(\nu\beta,\pm\omega_c\beta)=\frac{1}{2\pi}\left[-(\pm\omega_c+i\nu)\beta\Psi\left(1+i\frac{\nu\beta}{2\pi}\right)-(\pm\omega_c-i\nu)\beta\Psi\left(1-i\frac{\nu\beta}{2\pi}\right)\pm2\omega_c\beta\Psi\left(1+\frac{\pm\omega_c\beta}{2\pi}\right)\right],
\ee
with $\Psi(x)$ being the Digamma function. 
All together we obtain (changing the variable in the argument from $\nu$ to $\omega$),
\be\label{appC}
C_{OD}(\omega)= \frac{\pi}{\pm2\omega_c}\frac{\alpha}{\beta}\frac{\omega_c^2}{\omega_c^2 +\omega^2}\left[\omega_c^2 -\omega^2{\rm coth}(\omega\beta/2) +\frac{2\omega}{\beta}F(\omega\beta,\pm\omega_c\beta)\right].
\ee
 For $C(0) = \int_0^{\infty} d\omega \frac{J(\omega)}{\beta\omega} = \frac{Q}{\beta}$, we simply have
\be
C_{OD}(0)=\frac{Q_{OD}}{\beta} = \frac{\pi}{2}\frac{\alpha(\pm\omega_c)}{\beta}.
\ee

Then, $C_{OD}'(\omega)$ is "just" the derivative of $C_{OD}(\omega)$, which gives
\be\label{appdC}
C_{OD}'(\omega)= \frac{\pi}{2}\frac{\alpha}{\beta}\frac{\omega_c^2}{\omega_c^2 +\omega^2}\left[\frac{-2\omega\omega_c}{\omega_c^2+\omega^2}[1+{\rm coth}(\omega\beta/2)] -\frac{\omega^2\beta}{2\omega_c}{\rm coth'}(\omega\beta/2)+\frac{2}{\omega_c\beta}\frac{\omega_c^2-\omega^2}{\omega_c^2+\omega^2}F(\omega\beta,\omega_c\beta) +\frac{2\omega}{\omega_c\beta}F'(\omega\beta,\omega_c\beta)\right],
\ee
where ${\rm coth'}(x):=\frac{e^x}{{\rm sinh}x}(1-{\rm coth}x)$ is simply the derivative of ${\rm coth}(x)$, and 
\bea
F'(\omega\beta,\omega_c\beta)&:=&\frac{\partial}{\partial \omega} F(\omega\beta,\omega_c\beta)\nn\\
&=& \frac{\beta}{2\pi}\left[-i\Psi\left(1+i\frac{\omega\beta}{2\pi}\right)+i\Psi\left(1-i\frac{\omega\beta}{2\pi}\right)+(\omega-i\omega_c)\frac{\beta}{2\pi}\Psi'\left(1+i\frac{\omega\beta}{2\pi}\right) + (\omega+i\omega_c)\frac{\beta}{2\pi}\Psi'\left(1-i\frac{\omega\beta}{2\pi}\right)\right]\nn\\
\eea
 with $\Psi'$ is the derivative of the Digamma function. \\

Again, if $\omega_c$ is complex, we simply have 
\be
C_{OD}'(\omega)= \frac{\pi}{\pm2\omega_c}\frac{\alpha}{\beta}\frac{\omega_c^2}{\omega_c^2 +\omega^2}\left[\frac{-2\omega\omega_c^2}{\omega_c^2+\omega^2}[1+{\rm coth}(\omega\beta/2)] -\frac{\omega^2\beta}{2}{\rm coth'}(\omega\beta/2)+\frac{2}{\beta}\frac{\omega_c^2-\omega^2}{\omega_c^2+\omega^2}F(\omega\beta,\pm\omega_c\beta) +\frac{2\omega}{\beta}F'(\omega\beta,\pm\omega_c\beta)\right].
\ee
\end{widetext}

\subsubsection{Under-damped spectral density}\label{appUD}
We now consider an under-damped spectral density of the form $J_{UD}(\omega)$ \eqref{JUD},
\be
J_{UD}(\omega) := \omega\frac{2}{\pi}\frac{\gamma_{UD}\Omega^2\lambda^2}{(\Omega^2-\omega^2)^2+(\gamma_{UD}\Omega\omega)^2}.
\ee
 Such spectral densities can lead to difficulties related to analytical integration. This can be circumvented by expressing $J_{UD}$ as the difference of two over-damped spectral densities, 
\be
J_{UD}(\omega) = J_{OD}^{-}(\omega)-J_{OD}^{+}(\omega)
\ee
where
\be
J_{OD}^{\pm}(\omega):=\frac{2}{\pi}\frac{\gamma_{UD}\Omega^2\lambda^2}{\omega^2_{+}-\omega^2_{-}} \frac{\omega}{\omega^2+\omega_{\pm}^2},
\ee
with $\omega^2_{\pm}:= \Omega^2\left(\frac{\gamma_{UD}^2}{2}-1\pm\gamma_{UD}\sqrt{\frac{\gamma_{UD}^2}{4}-1}\right)$, always positive for $\gamma_{UD}\geq2$, and complex for $\gamma_{UD}<2$. Using this mapping, we straightforwardly obtain 
\bea\label{appcud}
C_{UD}(\omega) = C_{OD}^{-}(\omega)-C_{OD}^{+}(\omega),\nn\\
C'_{UD}(\omega) = {C'}_{OD}^{-}(\omega)-{C'}_{OD}^{+}(\omega),
\eea
where $C_{OD}^{\pm}(\omega)$ and ${C'}_{OD}^{\pm}(\omega)$ are given by the above expressions \eqref{appC} and \eqref{appdC} substituting $\omega_c^2$ by $\omega_{\pm}^2$, and $\alpha$ by $\alpha_{\pm}:= \frac{2}{\pi}\frac{\gamma_{UD}\Omega^2\lambda^2/\omega_{\pm}^2}{\omega^2_{+}-\omega^2_{-}}=\frac{\lambda^2}{\pi\Omega^2}f_{\pm}(\gamma_{UD})$ with 
\be
f_{\pm}(\gamma_{UD}) = \frac{1}{\left(\frac{\gamma_{UD}^2}{2}-1\pm\gamma_{UD}\sqrt{\frac{\gamma_{UD}^2}{4}-1}\right)\sqrt{\frac{\gamma_{UD}^2}{4}-1}}.
\ee
However, one has to be careful (see \eqref{cpxint}) for $\gamma_{UD}<2$ since $\omega_{\pm}^2$ becomes complex. One can verifies that for all $\gamma_{UD}>0$, $\Re \sqrt{\omega_{\pm}^2} >0$, so that the above expressions \eqref{appcud} still hold without change of signs, namely,
\bea
C_{UD}(\omega)&=& \frac{\pi}{2\omega_{-}}\frac{\alpha_{-}}{\beta}\frac{\omega_{-}^2}{\omega_{-}^2 +\omega^2}\Big[\omega_{-}^2 -\omega^2{\rm coth}(\omega\beta/2)\nn\\
&& \hspace{2.6cm}+\frac{2\omega}{\beta}F(\omega\beta,\omega_{-}\beta)\Big] \nn\\
&&- \frac{\pi}{2\omega_{+}}\frac{\alpha_{+}}{\beta}\frac{\omega_{+}^2}{\omega_{+}^2 +\omega^2}\Big[\omega_{+}^2 -\omega^2{\rm coth}(\omega\beta/2)\nn\\
&& \hspace{2.6cm}+\frac{2\omega}{\beta}F(\omega\beta,\omega_{+}\beta)\Big],
\eea
and similarly for $C'_{UD}(\omega)$. 

As a side note, we also show that the reorganisation energy is given by $Q_{UD} = \frac{\lambda^2}{\Omega}$.
From the above mapping into over-damped spectral densities, we have
\bea
Q_{UD} &=& Q_{OD, -} - Q_{OD,+}\nn\\
&=& \frac{\pi}{2} \alpha_{-}\sqrt{\omega_{-}^2} -\frac{\pi}{2} \alpha_{+}\sqrt{\omega_{+}^2} \nn\\
&=& \frac{\gamma_{UD}\Omega^2\lambda^2}{\omega_{+}^2-\omega_{-}^2}\left(\frac{1}{\omega_{-}}-\frac{1}{\omega_{+}}\right).\nn\\
\eea
Since,
\bea
\frac{1}{\omega^2_{+}-\omega^2_{-}}\left(\frac{1}{\omega_{-}}-\frac{1}{\omega_{+}}\right)&=&\frac{1}{\omega^2_{+}-\omega^2_{-}} \frac{\omega_{+}-\omega_{-}}{\omega_{+}\omega_{-}}\nn\\
&=&\frac{1}{\omega_{+}+\omega_{-}} \frac{1}{\omega_{+}\omega_{-}}\nn\\
&=& \frac{1}{\omega_{+}+\omega_{-}}\frac{1}{\Omega^2}\nn\\
&& \hspace{-3cm}= \frac{1}{\Omega^3}\Bigg[\left(\frac{\gamma_{UD}^2}{2}-1+\gamma_{UD}\sqrt{\frac{\gamma_{UD}^2}{4}-1}\right)^{1/2}\nn\\
&&\hspace{-2cm}+\left(\frac{\gamma_{UD}^2}{2}-1-\gamma_{UD}\sqrt{\frac{\gamma_{UD}^2}{4}-1}\right)^{1/2}\Bigg]^{-1}\nn\\
&& \hspace{-3cm}= \frac{1}{\Omega^3}\frac{1}{\gamma_{UD}}.
\eea
Note that one can easily see the last line by taking the square of what is in the square bracket.
Then, we finally obtain 
\be
Q_{UD} = \frac{\lambda^2}{\Omega},
\ee

Finally, note some useful identities with $\omega_{\pm}^2$,
\bea
&&\omega_{+}\omega_{-} = \Omega^2\nn\\
&& \omega_{+}+\omega_{-} = \Omega\gamma_{UD}.
\eea

\section{Some additional plots of $d_{\rm coh/pop}^{\rm ext}$ and criteria ${\rm cr}_i$}\label{addplots}
\begin{figure}
(a)\includegraphics[width=4cm, height=3cm]{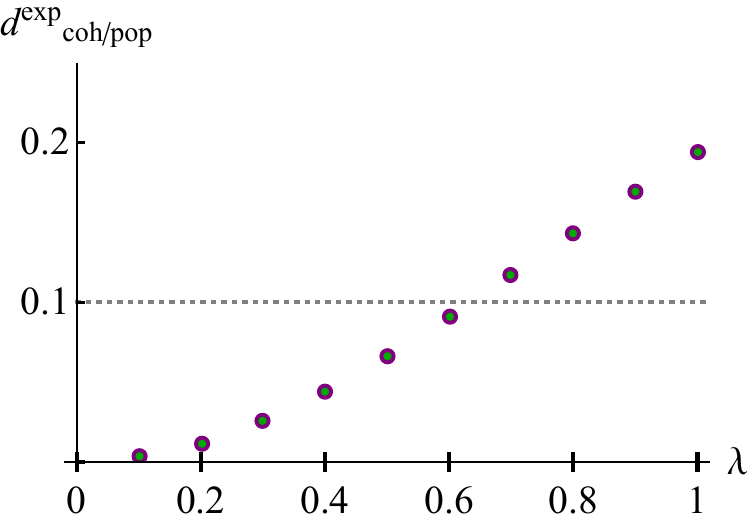}\includegraphics[width=4cm, height=3cm]{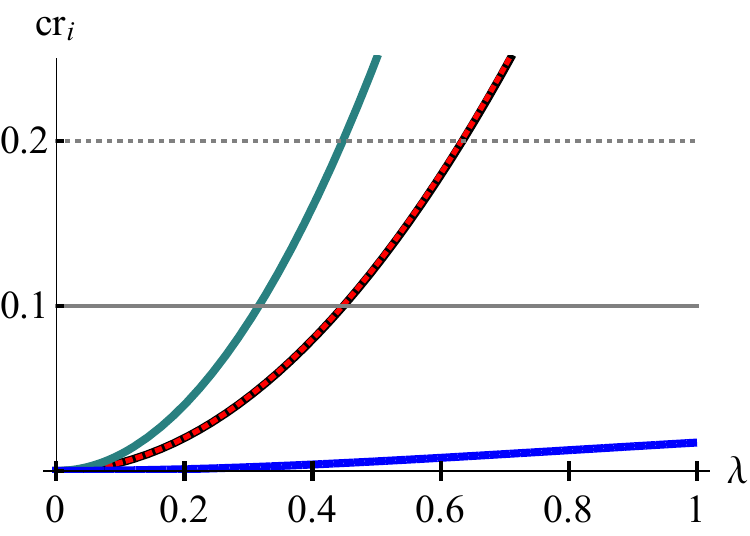}\\
(b)\includegraphics[width=4cm, height=3cm]{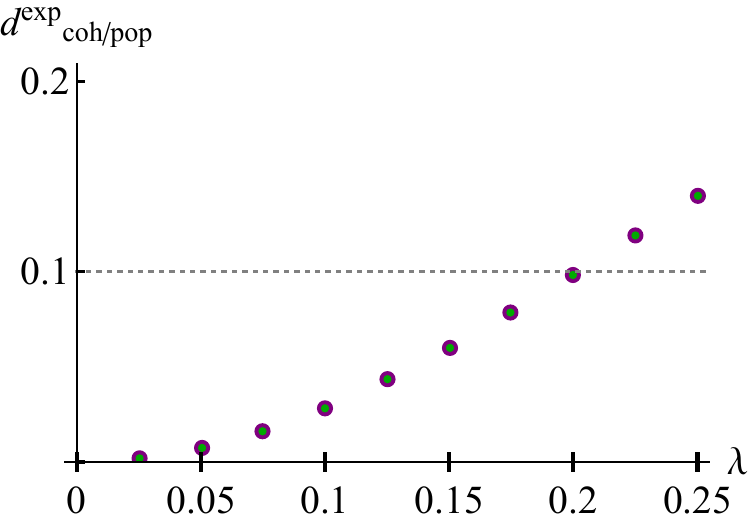}\includegraphics[width=4cm, height=3cm]{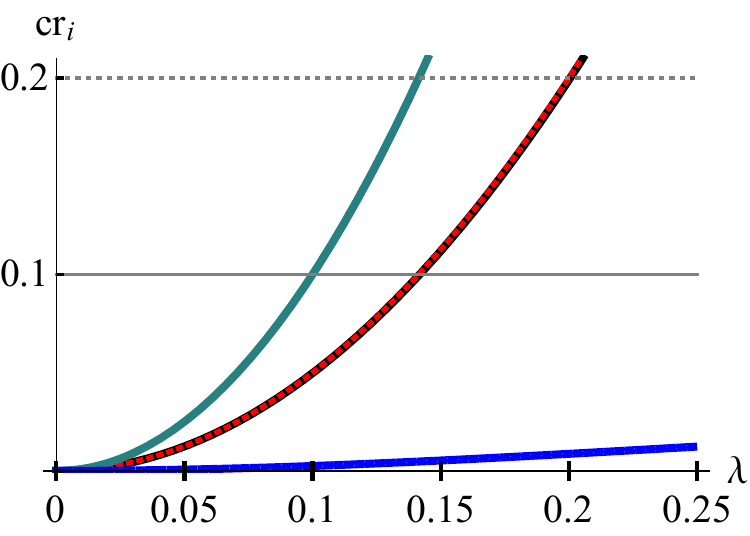}\\
(c)\includegraphics[width=4cm, height=3cm]{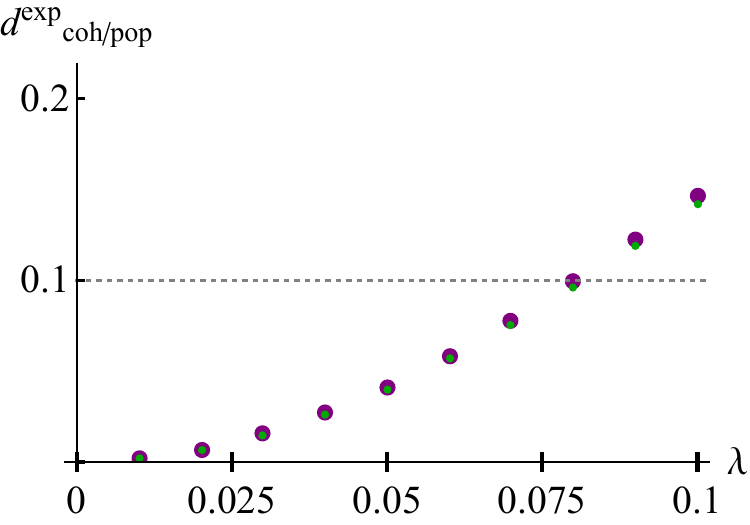}\includegraphics[width=4cm, height=3cm]{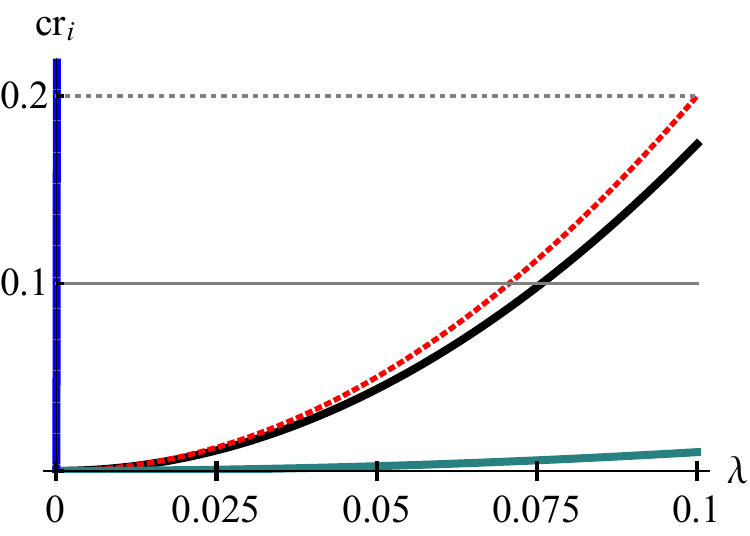}\\
(d)\includegraphics[width=4cm, height=3cm]{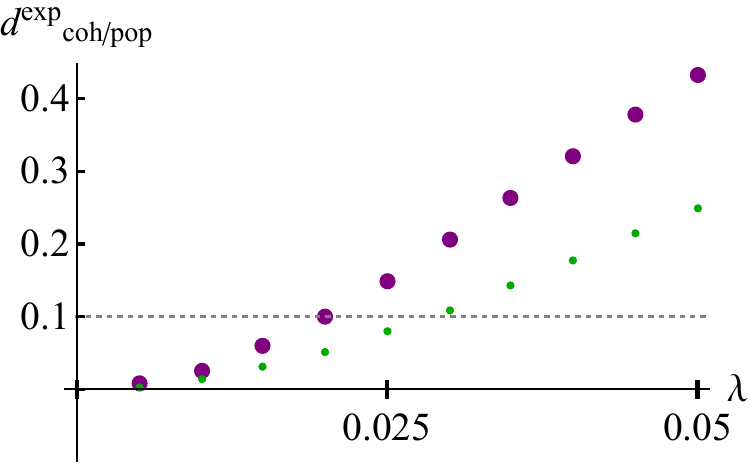}\includegraphics[width=4cm, height=3cm]{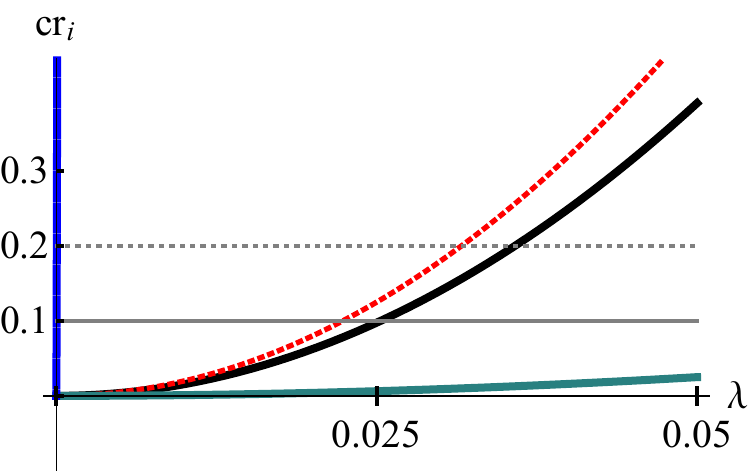}
\caption{On the left-hand-side, plots of $d_{\rm pop}^{\rm exp}$ (large purple dots) and $d_{\rm coh}^{\rm exp}$ (small green dots); on the right-hand-side plots of the criteria ${\rm cr}_i$, all in function of $\lambda$ (unit of $\omega_S$). The color convention for the criteria is the same as in the main text, namely, ${\rm cr}_1$ (black solid line), ${\rm cr}_2$ (blue solid line), ${\rm cr}_3$ (green solid line), and ${\rm cr}_4$ (red dashed line). The other parameters are as follows:(a) $\Omega/\omega_S=1$, $\omega_S\beta=0.5$; (b) $\Omega/\omega_S=0.1$, $\omega_S\beta=0.5$; (c) $\Omega/\omega_S=1$, $\omega_S\beta=20$; (d)$\Omega/\omega_S=0.1$, $\omega_S\beta=20$.  The remainder of the parameters are chosen as in figures of the main text, namely $\gamma_{UD} =0.1$, $r_z = \sqrt{0.75}$, $r=r_x+i r_y = 0.5$.} 
\label{dvsla}
\end{figure}

\begin{figure}[h]
(a)\includegraphics[width=4cm, height=3cm]{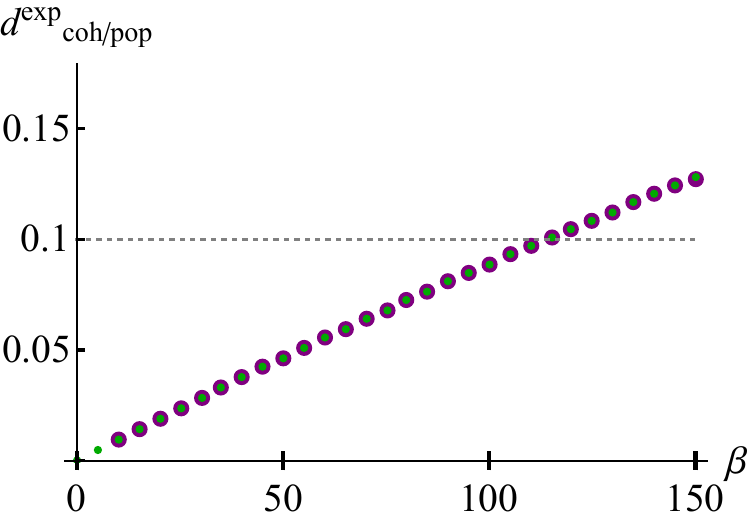}\includegraphics[width=4cm, height=3cm]{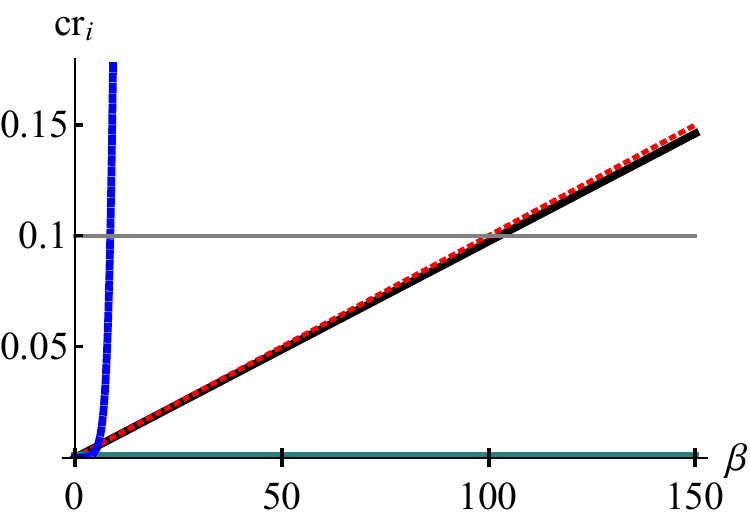}\\
(b)\includegraphics[width=4cm, height=3cm]{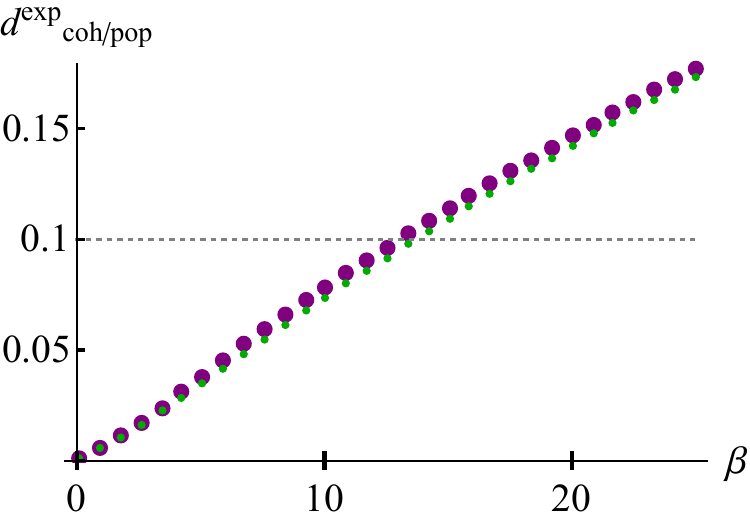}\includegraphics[width=4cm, height=3cm]{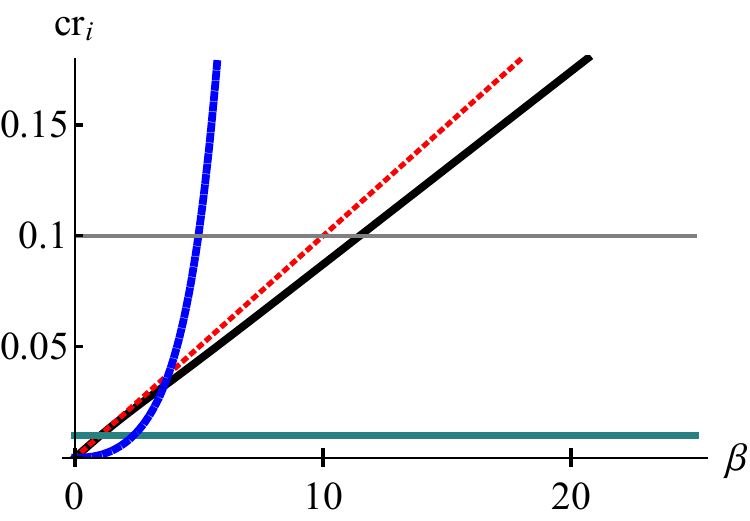}\\
(c)\includegraphics[width=4cm, height=3cm]{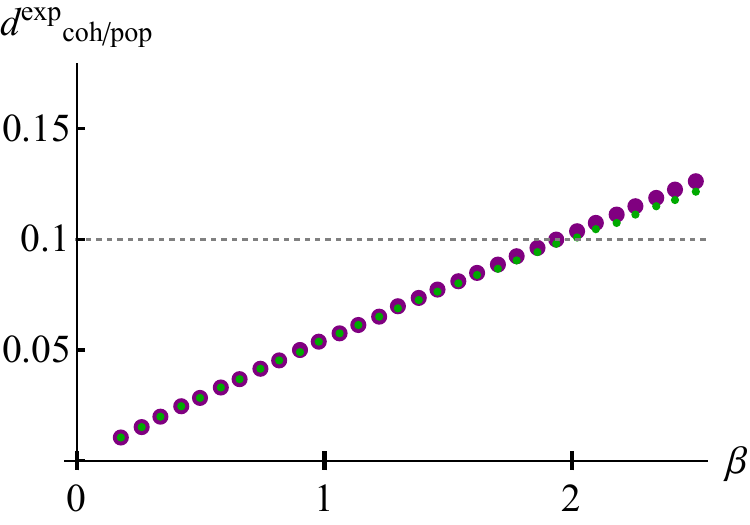}\includegraphics[width=4cm, height=3cm]{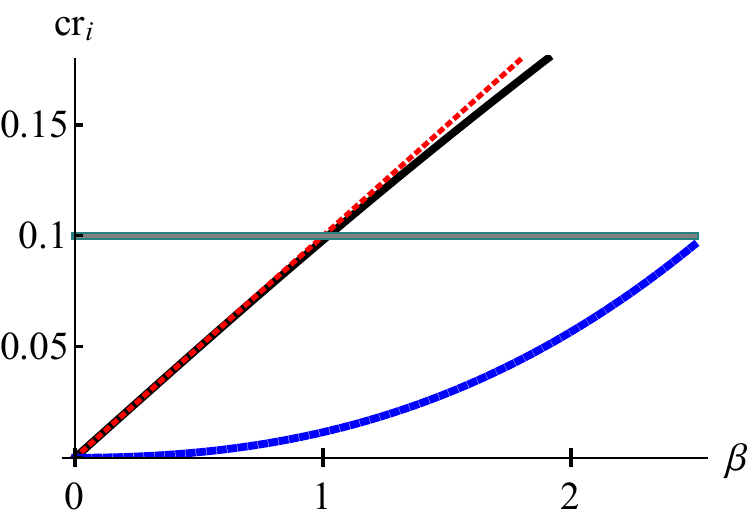}\\
(d)\includegraphics[width=4cm, height=3cm]{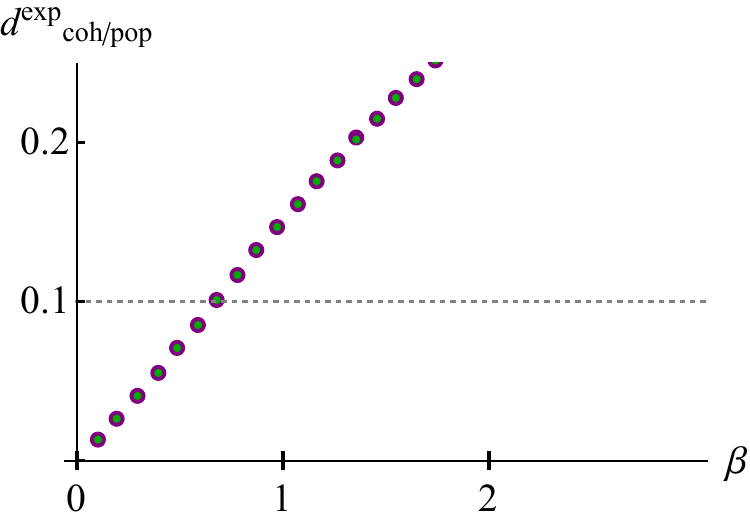}\includegraphics[width=4cm, height=3cm]{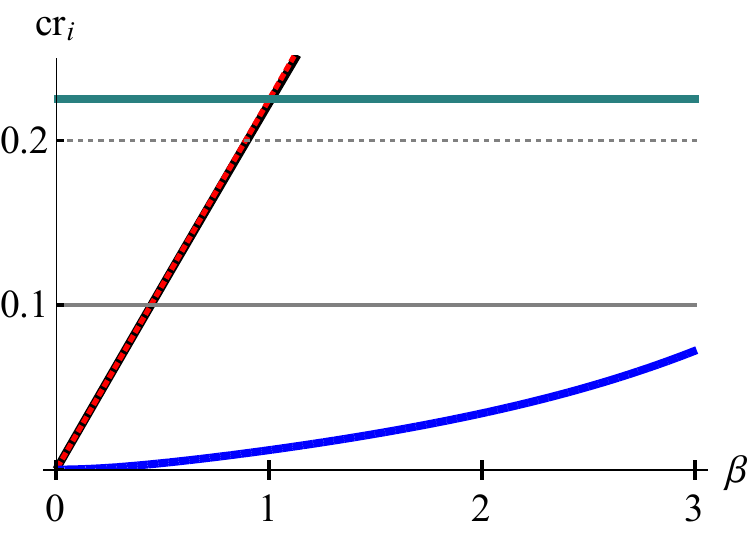}\\
(e)\includegraphics[width=4cm, height=3cm]{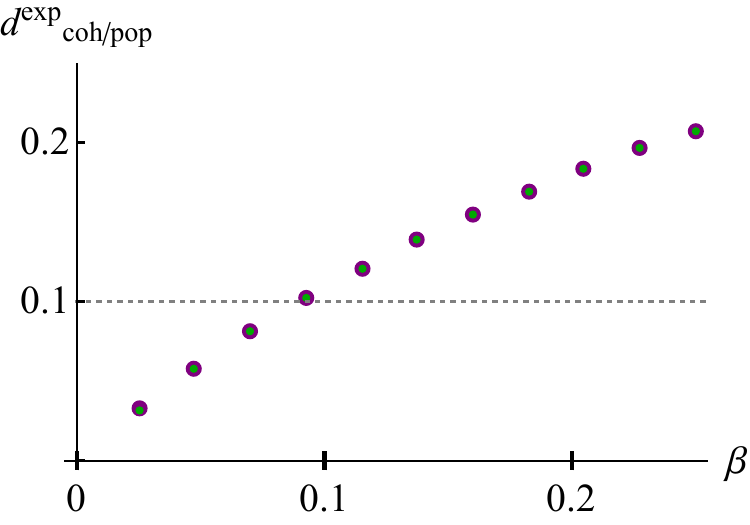}\includegraphics[width=4cm, height=3cm]{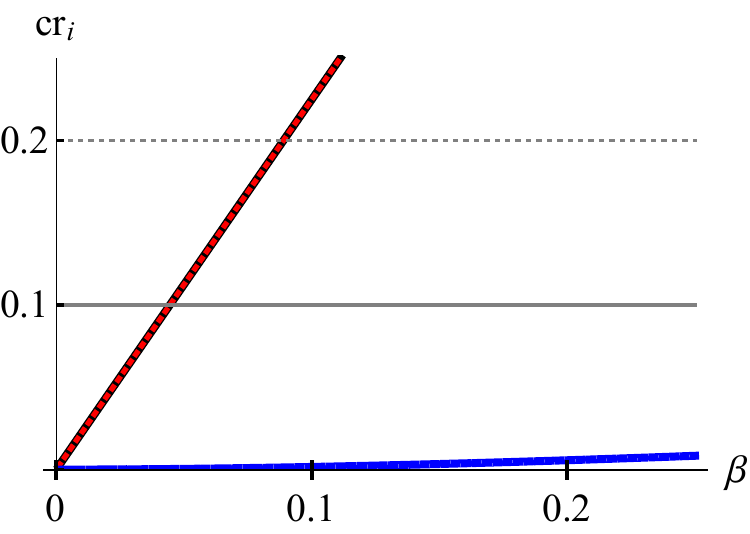}\\
\caption{On the left-hand-side, plots of $d_{\rm pop}^{\rm exp}$ (large purple dots) and $d_{\rm coh}^{\rm exp}$ (small green dots); on the right-hand-side plots of the criteria ${\rm cr}_i$, all in function of $\beta$ (unit of $\omega_S^{-1}$). The color convention for the criteria is the same as in the main text, namely, ${\rm cr}_1$ (black solid line), ${\rm cr}_2$ (blue solid line), ${\rm cr}_3$ (green solid line), and ${\rm cr}_4$ (red dashed line). The other parameters are as follows:(a) $\Omega/\omega_S=10$, $\lambda/\omega_S=0.1$; (b) $\Omega/\omega_S=1$, $\lambda/\omega_S=0.1$; (c) $\Omega/\omega_S=0.1$, $\lambda/\omega_S=0.1$; (d)$\Omega/\omega_S=10$, $\lambda/\omega_S=1.5$; (e)$\Omega/\omega_S=1$, $\lambda/\omega_S=1.5$. The remainder of the parameters are chosen as in figures of the main text, namely $\gamma_{UD} =0.1$, $r_z = \sqrt{0.75}$, $r=r_x+i r_y = 0.5$.} 
\label{dvsb}
\end{figure}

In this section we provide some additional plots in Figs. \ref{dvsla} and \ref{dvsb}, showing unambiguously that the criterion ${\rm cr}_4 = \beta Q$ (as well as ${\rm cr}_1$) accurately predicts the validity range of the perturbative expansion.

\end{document}